\documentclass[fleqn,usenatbib]{mnras}

\usepackage{newtxtext,newtxmath}
\usepackage[T1]{fontenc}

\DeclareRobustCommand{\VAN}[3]{#2}
\let\VANthebibliography\thebibliography
\def\thebibliography{\DeclareRobustCommand{\VAN}[3]{##3}\VANthebibliography}

\usepackage{float}
\usepackage{graphicx}	% Including figure files
\usepackage{amsmath}	% Advanced maths commands
\usepackage{xcolor}

\newcommand{\revision}[1]{{#1}\ignorespaces}
\title[Evaluating spurious correlations in PTA datasets]{Evaluating the prevalence of spurious correlations in pulsar timing array datasets}
\author[A. Zic et al.]{Andrew Zic,$^{1, 2}$\thanks{E-mail: andrew.zic@mq.edu.au}
George Hobbs,$^{2}$
R. M. Shannon,$^{3,4}$
Daniel Reardon,$^{3,4}$
Boris Goncharov,$^{5,6}$
\newauthor
N.~D.~Ramesh Bhat,$^{7}$
Andrew Cameron,$^{3,4}$
Shi Dai,$^{8,9,2}$
J.~R.~Dawson,$^{1, 2}$
Matthew Kerr,$^{10}$
\newauthor
R.~N.~Manchester,$^{2}$
Rami Mandow,$^{1, 2}$
Tommy Marshman,$^{1, 2}$
Christopher J. Russell,$^{11}$
\newauthor
Nithyanandan Thyagarajan,$^{12}$
X.~-J. Zhu$^{13}$
\\
$^{1}$School of Mathematical and Physical Sciences, and Research Centre in Astronomy, Astrophysics and Astrophotonics, Macquarie University, NSW 2109, Australia\\
$^{2}$Australia Telescope National Facility, CSIRO, Space and Astronomy, PO Box 76, Epping, NSW 1710, Australia\\
$^{3}$Centre for Astrophysics and Supercomputing, Swinburne University of Technology, PO Box 218, Hawthorn, VIC 3122, Australia\\
$^{4}$OzGrav: The ARC Centre of Excellence for Gravitational Wave Discovery, Australia\\
$^{5}$Gran Sasso Science Institute (GSSI), I-67100 L'Aquila, Italy\\
$^{6}$INFN, Laboratori Nazionali del Gran Sasso, I-67100 Assergi, Italy\\
$^{7}$International Centre for Radio Astronomy Research (ICRAR), Curtin University, Bentley, WA 6102, Australia\\
$^{8}$School of Science, Western Sydney University, Locked Bag 1797, Penrith, NSW 2751, Australia\\
$^{9}$National Astronomical Observatories, Chinese Academy of Sciences, Beijing 100101, China\\
$^{10}$Space Science Division, Naval Research Laboratory, Washington, DC 20375-5352, USA\\
$^{11}$CSIRO Scientific Computing, Australian Technology Park, Locked Bag 9013, Alexandria, NSW 1435, Australia\\
$^{12}$Australia Telescope National Facility, CSIRO, Space and Astronomy, PO Box 1130, Bentley, WA 6102, Australia\\
$^{13}$Advanced Institute of Natural Sciences, Beijing Normal University, Zhuhai 519087, China\\
}

\date{Accepted 2022 July 19. Received 2022 July 4; in original form 2022 May 4}

\pubyear{2022}

\begin{document}
\label{firstpage}
\pagerange{\pageref{firstpage}--\pageref{lastpage}}
\maketitle

\begin{abstract}
%CONTEXT
Pulsar timing array collaborations have recently reported evidence for a noise process with a common spectrum among the millisecond pulsars in the arrays. The spectral properties of this common-noise process are consistent with expectations for an isotropic gravitational-wave background (GWB) from inspiralling supermassive black-hole binaries. However, recent simulation analyses based on Parkes Pulsar Timing Array data indicate that such a detection may arise spuriously. %, and may represent the emergence of the first observational evidence for this phenomenon.
%HOW
In this paper, we use simulated pulsar timing array datasets to further test the robustness of the inference methods for spectral and spatial correlations from a GWB. 
%WHAT
%"cvfwsae bv" -- Singen
Expanding on our previous results, we find strong support (Bayes factors exceeding $10^5$) for the presence of a common-spectrum noise process in datasets where no common process is present, under a wide range of timing noise prescriptions per pulsar. We show that these results are highly sensitive to the choice of Bayesian priors on timing noise parameters, with priors that more closely match the injected distributions of timing noise parameters resulting in diminished support for a common-spectrum noise process. 
These results emphasize shortcomings in current methods for inferring the presence of a common-spectrum process, and imply that the detection of a common process is not a reliable precursor to detection of the GWB. Future searches for the nanohertz GWB should remain focussed on detecting spatial correlations, and make use of more tailored specifications for a common-spectrum noise process. 
\end{abstract}


%%%%%%%%%%%%%%%%%%%%%%%%%%%%%%%%%%%%%%%%%%%%%%%%%%

%%%%%%%%%%%%%%%%% BODY OF PAPER %%%%%%%%%%%%%%%%%%

\section{Introduction}
Pulsar timing arrays (PTAs) consist of sets of millisecond pulsars (MSPs) exhibiting high timing stability \citep{1990ApJ...361..300F}. Among myriad scientific goals \citep{2013PASA...30...17M}, the primary aim of PTA experiments is the detection and characterization of the isotropic stochastic gravitational-wave background \citep[GWB; e.g.][]{2005ApJ...625L.123J}. Current PTA experiments include the European PTA \citep[EPTA;][]{2013CQGra..30v4009K}, the North American Nanohertz Observatory for Gravitational waves \citep[NANOGrav;][]{2013CQGra..30v4008M}, the Parkes PTA \citep[PPTA;][]{2013PASA...30...17M}, the Indian PTA \citep[InPTA;]{2018JApA...39...51J}, which all comprise the International PTA \citep[IPTA;][]{2010CQGra..27h4013H}. Other nascent PTA collaborations, such as the Chinese PTA \citep{2016ASPC..502...19L}, and experiments with the MeerKAT telescope in South Africa \citep[e.g.][]{2020PASA...37...28B} may join efforts with the IPTA in coming years. 

Some PTA collaborations have recently detected a noise process with spectral properties that appear common among all pulsars, possibly representing the emergence of the GWB signal in their datasets \citep[e.g.][]{2020ApJ...905L..34A}. However, because these datasets are currently in a sub-threshold and highly model-dependent regime for GWB detection \citep{2020ApJ...905L...6H,2021MNRAS.502..478G,2021PhRvD.103f3027R,2021ApJ...911L..34P}, these findings require robust validation checks in order to understand their significance. This forms the underlying motivation behind this paper.

The largest contribution to the GWB is expected to come from a cosmological population of supermassive black-hole binaries \citep[SMBHBs; ][]{2015MNRAS.451.2417R,2013MNRAS.433L...1S,2003ApJ...590..691W}, but other more exotic processes such as cosmological phase transitions \citep{2021PhRvL.127y1303X, 2021PhRvL.127y1302A, 2017EPJC...77..570K}, vibration of cosmic strings \citep{2010PhRvD..81j4028O}, and quantum fluctuations in the early universe \citep{2016PhRvX...6a1035L,1982PhLB..108..389L, 1980PhLB...91...99S,1976JETPL..23..293G} are also expected to contribute.

For a cosmological population of SMBHBs in circular orbits, with energy loss dominated by gravitational-wave (GW) emission within the PTA band, the GW strain is \citep{2001astro.ph..8028P}
\begin{equation}
    h_c(f) = A_\text{GWB} \left(\frac{f}{1\,\mathrm{yr}^{-1}}\right)^{\alpha}~\text{,}
\end{equation}
where $A_\text{GWB}$ is the gravitational-wave strain amplitude, and $\alpha = -2/3$ is the strain spectral index.

The GWB strain at the Earth and at the pulsar will cause stochastic fluctuations in the pulse times of arrival (ToAs) on timescales of years to decades. The GWB therefore produces a red noise signal\footnote{A temporally correlated noise process with higher spectral power at lower fluctuation frequencies compared to higher frequencies.} in the timing residuals\footnote{Timing residuals are formed by subtracting a model for the pulse ToAs from the measured ToAs}. The GWB-induced red noise has a cross-correlated power spectral density (PSD) across pulsar pairs given by
\begin{equation}
    \label{eq:ccpower_spec}
    \mathcal{P}_{ab}(f| A_\text{GWB}, \gamma) = \Gamma_{ab} (\zeta) \frac{A_\text{GWB}^2}{12 \pi^2} \left(\frac{f}{1\,\mathrm{yr}^{-1}}\right)^{-\gamma}~\text{.}
\end{equation}
Here, $\gamma = 3 -2\alpha$ is the spectral index, equal to $13/3$ in the case of an isotropic GWB from inspiralling SMBHBs, and $\Gamma_{ab}(\zeta)$ is the overlap reduction function (ORF), which describes the spatial correlation of the GWB signal between distinct pulsars $a$ and $b$ separated by an angle $\zeta$. For an isotropic GWB from inspiralling circular SMBHBs, the ORF is known as the Hellings-Downs (HD) correlation function \citep{1983ApJ...265L..39H}:
\begin{equation}
\label{eq:HD}
    \Gamma_{ab}(\zeta) = \frac{1}{2} - \frac{1}{4}x + \frac{3}{2}x\ln(x)~\text{,} 
\end{equation}
where $x = (1 - \cos \zeta)/2$. The isotropic nature of the GWB implies that the induced timing fluctuations can be described by the same spectrum (as in Eq. \ref{eq:ccpower_spec}) among all pulsars and the temporal correlation of timing residuals among the PTA pulsars will vary spatially according to Eq. \ref{eq:HD}. 

PTA collaborations view the detection of HD correlations (as in Eq. \ref{eq:HD}) as unambiguous evidence for the nanohertz GWB. 
However, as well as the spatially correlated component, the GWB signal also contains an uncorrelated component, with the former arising from the strain at the Earth (``Earth term'') and the latter at each pulsar (``pulsar term''). 
Prior to detection of HD spatial correlations, it is possible that a GWB signal may first emerge in the autocorrelation terms of the full inter-pulsar correlation matrix \citep{2021PhRvD.103f3027R,2021ApJ...911L..34P}, particularly for PTAs with a modest number of pulsars ($\lesssim 50$).
The autocorrelations are sensitive both to the Earth and pulsar term, while the cross-correlations are sensitive primarily to the Earth term alone \citep{2018JPhCo...2j5002M}. As in Equation \ref{eq:ccpower_spec}, the cross-correlated PSD is modulated by $\Gamma_{ab}(\zeta)$, \revision{which has a weighted average of $\sim 0.016$ and standard deviation $0.026$ for the angular separations of PPTA pulsars}. This implies that the cross-correlated PSD amplitude is significantly attenuated relative to the autocorrelated PSD amplitude. In the autocorrelation terms, the GWB is expected to manifest as a common-spectrum red noise process -- i.e., a red noise process that is present within all pulsar timing residuals, described by a single PSD:
\begin{equation}
    \label{eq:rednoise_spec}
    \mathcal{P}(f| A, \gamma) = \frac{A^2}{12 \pi^2} \left(\frac{f}{1\,\mathrm{yr}^{-1}}\right)^{-\gamma}~\text{.} 
\end{equation}
In the literature, this common-spectrum process is usually termed the ``common red noise'' (CRN), which we will also use for the remainder of this paper. 

Currently, none of the PTA collaborations has reported significant evidence for HD-correlated signals in their datasets. Recently, however, NANOGrav, the PPTA, the EPTA, and the IPTA have reported strong evidence for the presence of a CRN process in recent data releases \citep{2020ApJ...905L..34A, 2021ApJ...917L..19G, 2021MNRAS.508.4970C, 2022MNRAS.510.4873A}. Though there is some variance among the best estimates, the reported CRN properties are consistent within uncertainties. However, the inferred amplitudes for the CRN at a fixed spectral index of $\gamma = 13/3$, ranging from $\sim 2.0\times 10^{-15}$ to $3.0\times 10^{-15}$ at a reference frequency of $1\,\text{yr}^{-1}$, are in tension with previously set 95\,per\,cent credible interval upper limits from NANOGrav \citep[ $ A_\text{GWB} < 1.45\times 10^{-15}$;][]{2018ApJ...859...47A} and the PPTA \citep[$A_\text{GWB}<1.0\times 10^{-15}$;][]{2015Sci...349.1522S}.

%NG: 5.52
%PPTA: 4.11 +0.52-0.41
%EPTA: 3.78 +0.69-0.59
%IPTA: 4.0 \pm 0.9

These discrepancies have been a point of concern among PTA collaborations
\citep[e.g.][]{2020ApJ...905L..34A}, but recent work by \citet{2022ApJ...932..105J} suggests that upper limits are more likely to be under-estimated when they are formed using a subset of pulsars from a PTA, as in \citet{2015Sci...349.1522S}. While this offers a possible explanation for the discrepancies between CRN properties and previous upper limits, investigations by \citet{2021ApJ...917L..19G} have found that a CRN can be ``detected'' in simulated datasets only containing individual pulsar noise terms with disparate characteristics. This raises concerns that a CRN signal can be strongly influenced by, or arise entirely from, independent pulsar noise processes that have no relationship with the GWB.

One of the most important noise processes present in individual pulsars is ``timing noise'' \citep{1975ApJS...29..453G,1999ptgr.conf..141L,2010MNRAS.402.1027H, 2013CQGra..30v4002C,2019MNRAS.489.3810P}, also known as ``spin noise''  -- stochastic, time-correlated variations in the pulsar ToAs thought to be driven by rotational irregularities and other pulsar-intrinsic fluctuations. Concerns about inferences on the CRN pr ocess have arisen on the basis that some MSPs may exhibit similar timing noise characteristics \citep{2010ApJ...725.1607S, 2021MNRAS.502..478G}, which may result in a false-alarm detection of a CRN. Indeed, \citet{2021MNRAS.502.3113M} suggest that pulsar timing noise induced by spin irregularities has a spectral index of $4$, close to the $13/3$ value expected for a GWB. Incorrect or incomplete models for pulsar-intrinsic noise terms can bias searches for, or prevent detection of, the GWB \citep{2021ApJ...917L..19G,2020ApJ...905L...6H, 2015MNRAS.449.3293L, 2013CQGra..30v4002C}. 

  It is possible that the detection of a CRN process among PTA collaborations is truly the first emerging evidence of the GWB \citep{2021PhRvD.103f3027R,2021ApJ...911L..34P}. However, the tension with previous upper limits \citep{2022ApJ...932..105J}, and false detections in simulations presented by \citet{2021ApJ...917L..19G}, and in this work, warrant further investigation into present methodologies and biases involved in detecting a CRN. There have been recent efforts to improve inference methods for the CRN \citep[e.g.][]{2022ApJ...932L..22G}. As we will show in this work, developments such as these are necessary to \revision{consolidate recent detections of a CRN as milestones toward the detection of the GWB via spatial correlations}.

While understanding the subtleties involved in CRN inference is important in the context of recent results, the key to unambiguously detecting the GWB lies in the spatial correlations. Therefore, understanding the robustness of spatial correlation inference techniques under different contexts is critical \citep{2016MNRAS.455.4339T, 2017PhRvD..95d2002T}. Pulsar timing noise is expected to be one of the main obstacles to GWB detection \citep{2013CQGra..30v4002C, 2015MNRAS.449.3293L, 2017PhRvD..95d2002T}, understanding its influence on spatial correlation inferences is particularly pertinent.

In this paper, we use simulated pulsar timing array datasets containing timing noise to explore the biases in current techniques for inferring the presence of common-spectrum and spatially correlated signals in pulsar timing array datasets . In Section \ref{sec:simulations}, we describe the simulation process and present the properties of the simulated datasets. In Section \ref{sec:analysis}, we present our analysis of these simulations, and discuss implications for recent pulsar timing array results and methodologies. Section \ref{sec:Discussion} contains concluding discussion and remarks for this work.

\section{Simulations}
\label{sec:simulations}
% \subsection{PTASimulate description}
We constructed pulsar timing array datasets using \textsc{PTASimulate}\footnote{\url{https://bitbucket.org/psrsoft/ptasimulate}}, a package for simulating pulsar ephemerides and ToAs in a format suitable for \textsc{tempo2}~\citep{2006MNRAS.372.1549E}. The package can be used to inject various stochastic and deterministic signals into the ToAs, which can then be used for studying observing strategies, telescope and PTA sensitivities, GWB analysis techniques, and many other topics relevant to PTA datasets.

While \textsc{PTASimulate} can simulate realistic datasets with a wide range of pulsar timing phenomena, 
in this work we chose to simulate ToAs recorded at a regular cadence and at a single frequency band, with uniform ToA uncertainties per pulsar, chosen between 90 to 500\,ns based on similarities to PPTA datasets. We made these choices primarily to reduce computational costs while exploring a wide parameter space, but also to minimize dataset complexity that could obfuscate interpretation of our analysis.

Following \citet{2021ApJ...917L..19G}, we simulated timing residuals for the 26 pulsars in the PPTA second data release \citep[DR2; ][]{2020PASA...37...20K}, with a regular cadence of 40 days, and over a time-span of 20 years. For each pulsar, we injected timing noise as a red noise signal with a power-law PSD $\mathcal{P}_\text{TN}$ parametrized as
\begin{equation}
\label{eq:ptasim_psd}
    \mathcal{P}_\text{TN} (f | P_0, \gamma, f_c ) = \frac{P_0}{\left(1 + \left(\frac{f}{f_c}\right)^{2}\right)^{\gamma /2 }}~[\text{s}^{3}],
\end{equation}
where $f$, $f_c$ are the fluctuation and corner frequencies respectively in units of $\text{yr}\,^{-1}$, $P_0$ is the PSD amplitude, and $\gamma$ is the spectral index. \revision{This parametrization explictly encodes a low-frequency ``corner'' in the PSD, below which the PSD plateaus at a constant value. In the limit $f \gg f_c$, Eq. \ref{eq:ptasim_psd} simplifies to a standard power-law parametrization as in Eq. \ref{eq:rednoise_spec}, with $P_0 = A^{2} f_c^{\gamma} / (12 \pi^2)$}.

For each pulsar, the PSD amplitude was drawn from a log-uniform distribution with a median value $P_{0,m} = 10^{-23}$ (corresponding to $\log_{10} A_m = -14.46$) and a width $\Delta \log_{10} P_m$, and the spectral index was drawn from a uniform distribution with a median value $\gamma_m = 4$ and a width $\Delta \gamma_m$.\footnote{The subscript $m$ indicates the median value of the distribution of injected timing noise parameters.} \revision{The choice of these median values was made to approximately match the characteristics of the CRN recently detected in the PPTA DR2 \citep[e.g.][]{2021ApJ...917L..19G}, so that we could investigate the robustness of CRN detections in similar datasets}. In \citet{2021ApJ...917L..19G}, we simulated datasets where the full width of the input uniform distribution for the PSD amplitude (hereafter termed $\Delta \log_{10} P_0$) and spectral index (hereafter termed $\Delta \gamma$) was increased simultaneously (i.e., we only explored a one-dimensional path in the $(\Delta \log_{10} P_0, \Delta \gamma)$ parameter space). In this work, we extended this analysis by exploring the $(\Delta \log_{10} P_0, \Delta \gamma)$ parameter space in both dimensions. We sampled the $(\Delta \log_{10} P_0, \Delta \gamma)$ parameter space in a regularly-spaced $11\times 11$ grid, where $\Delta \log_{10} P_0$ varied from $0$ to $14$, and $\Delta \gamma$ varied from $0$ to $8$, around the central values $\log_{10} P_{0,m} = -23$ ($\log_{10} A_m = -14.46$) and $\gamma_m = 4.0$. In this description, $(\Delta \log_{10} P_0, \Delta \gamma) = (0, 0)$ corresponds to a true common-spectrum red noise process, and increases of $\Delta \log_{10} P_0$ and $\Delta \gamma$ correspond to increasingly disparate pulsar timing noise properties.

For each choice of $\Delta \log_{10} P_0$ and $\Delta \gamma$, we drew a PSD amplitude $P_0$ and spectral index $\gamma$ for each pulsar, and simulated 100 realizations of timing noise according to these chosen parameters. Altogether with 121 choices of timing noise parameters, with 100 realizations each, this resulted in 12,100 simulated datasets. The large number of realizations enabled us to explore detection statistics and perform false alarm analyses across the $(\Delta \log_{10} P_0, \Delta \gamma)$ parameter space. We note that the values of $\log_{10} P_0$ and $\gamma$ for each pulsar were drawn independently from the uniform distribution with widths $\Delta \log_{10} P_0$, $\Delta \gamma$, with no regard given to the properties of any particular pulsar in real datasets.

\begin{figure}
    \centering
    \includegraphics[width=\linewidth]{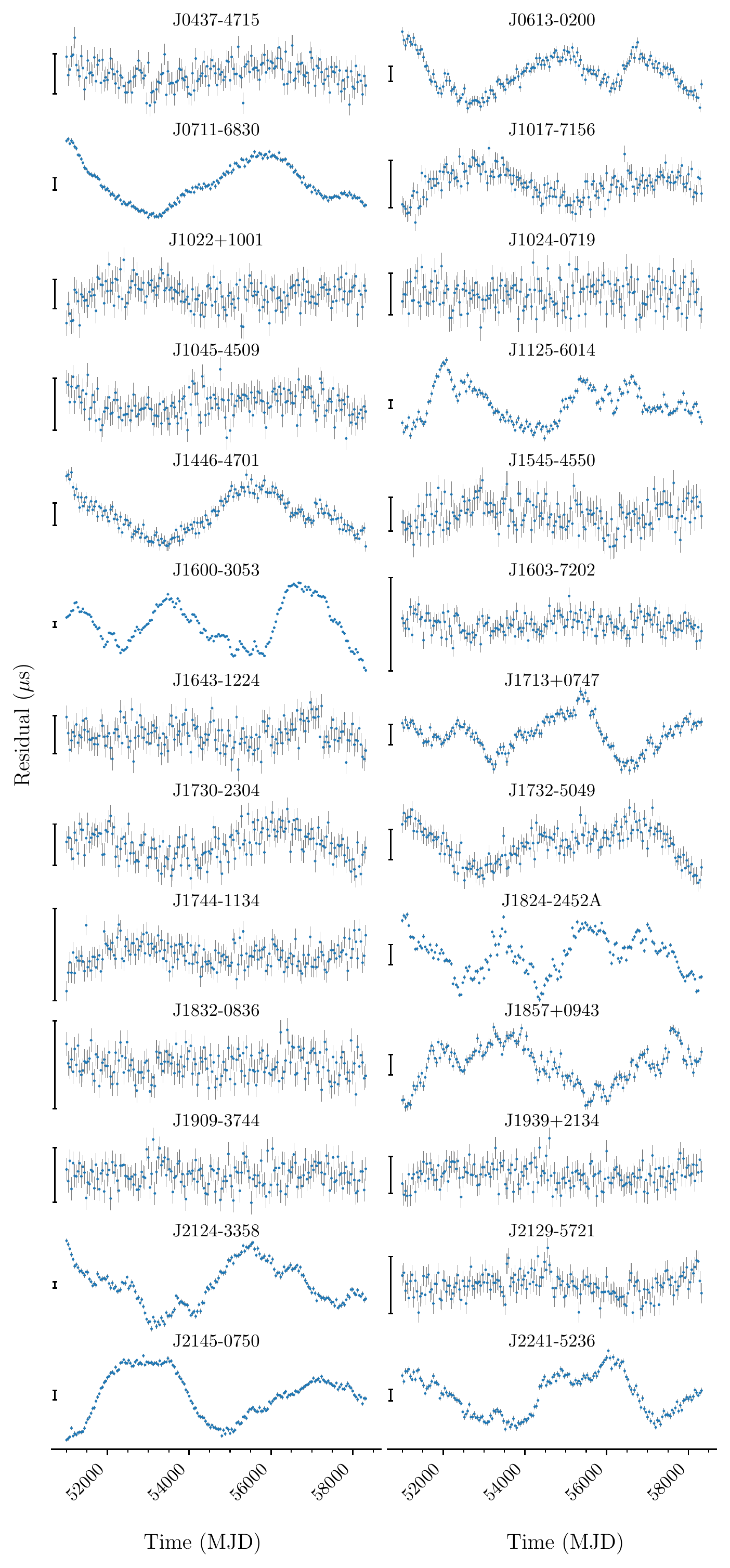}
    \caption{Timing residuals for a simulated dataset with $\Delta \log_{10} P_0 = 2.80$, $\Delta \gamma = 4.00$. We show a $1\, \mu s$ scale bar in the left of each sub-figure for reference. Note that the simulated residuals are not necessarily reflective of the residuals for each pulsar in real datasets.}
    \label{fig:residuals}
\end{figure}

\begin{figure}
    \centering
    \includegraphics[width=\linewidth]{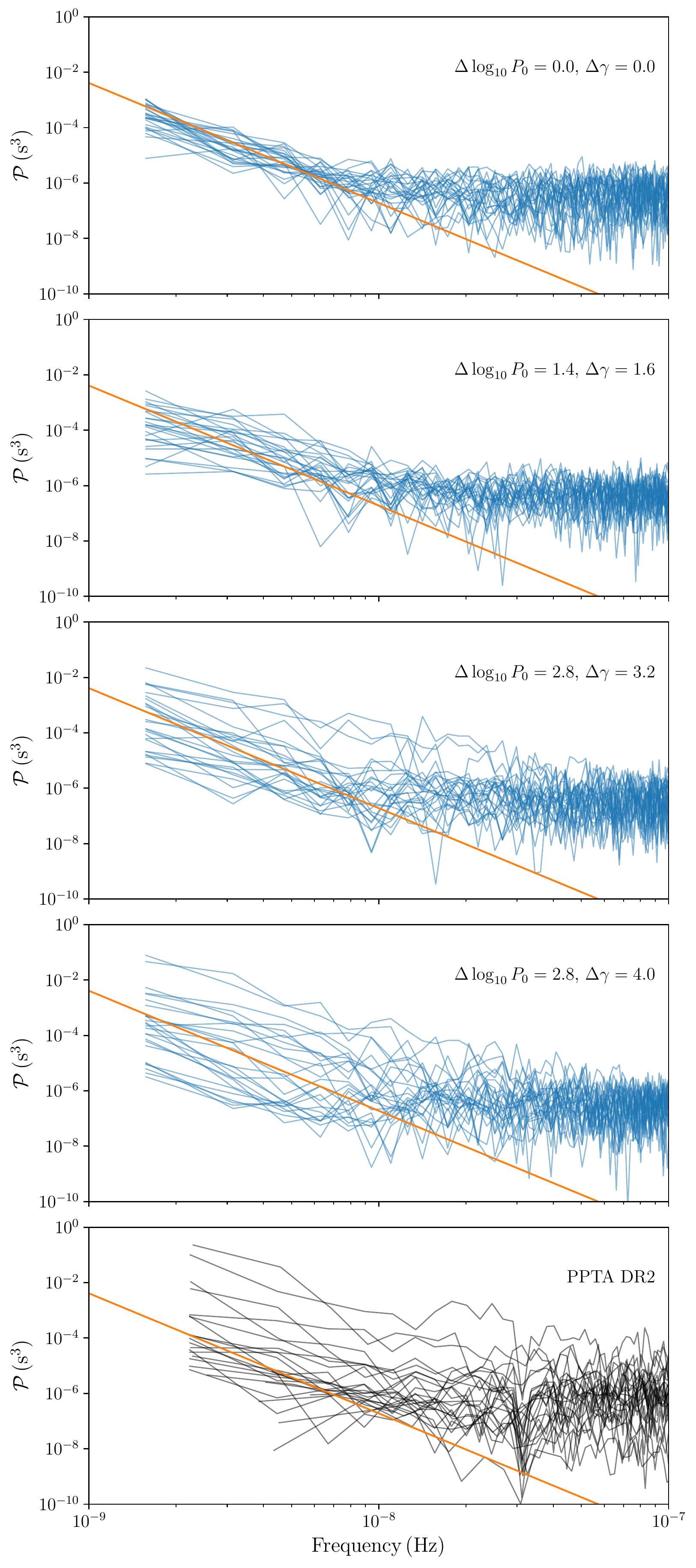}
    \caption{Power spectral densities of timing residuals from four simulated datasets, each with timing noise sampled from distributions of different widths (rows 1 to 4, coloured blue) and for the PPTA DR2 dataset (row 5, coloured black). The spectra have been calculated using a generalized least-squares technique \citep{2011MNRAS.418..561C}. Distribution widths for the simulated input timing noise parameters, $(\Delta \log_{10} P_0, \Delta \gamma)$ are $(0, 0)$, $(1.4, 1.6)$, $(2.8, 3.2)$, and $(2.8, 4.0)$, from top toward bottom panels. We also show a reference spectrum (orange) in each panel, with $A = 2.2\times 10^{-15}$, $\gamma = 13/3$, representing the CP2 model found in PPTA DR2 \citep{2021ApJ...917L..19G}.}
    \label{fig:noise_PSDs}
\end{figure}

In Figure \ref{fig:residuals} we show timing residuals in one realization of a simulation with $\Delta \log_{10} P_0= 2.8$,  $\Delta \gamma = 4.0$, and in Figure \ref{fig:noise_PSDs}, we show the PSDs for the 26 pulsars, with increasing degrees of variation in the injected timing noise parameters.

We do not inject any GWB or other spatially-correlated signals into our simulations. While the GWB signal likely exists within real datasets, even if it is low in amplitude \citep{2018MNRAS.477.2599B,2017MNRAS.470.4547D,2016ApJ...819L...6T,2015Sci...349.1522S}, our analysis on datasets containing only independent pulsar timing noise terms allows us to test current methodologies in the ``worst-case'' scenario of PTA datasets dominated by pulsar noise terms. By doing this, we aim to investigate the extent that recent detections of a CRN \citep{2020ApJ...905L..34A, 2021ApJ...917L..19G,2021MNRAS.508.4970C, 2022MNRAS.510.4873A} could be influenced by timing noise. Furthermore, there have already been several detailed GWB injection-recovery analyses presented elsewhere, so we do not repeat those analyses here \citep[e.g.,][]{2022PhRvD.105h4049T, 2020ApJ...905L...6H, 2018PhRvD..98d4003V,2016MNRAS.455.4339T}. 

\section{Common-spectrum and spatial correlation analysis}%Discussion}
\label{sec:analysis}

\begin{figure}
    \centering
    \includegraphics[width=\linewidth]{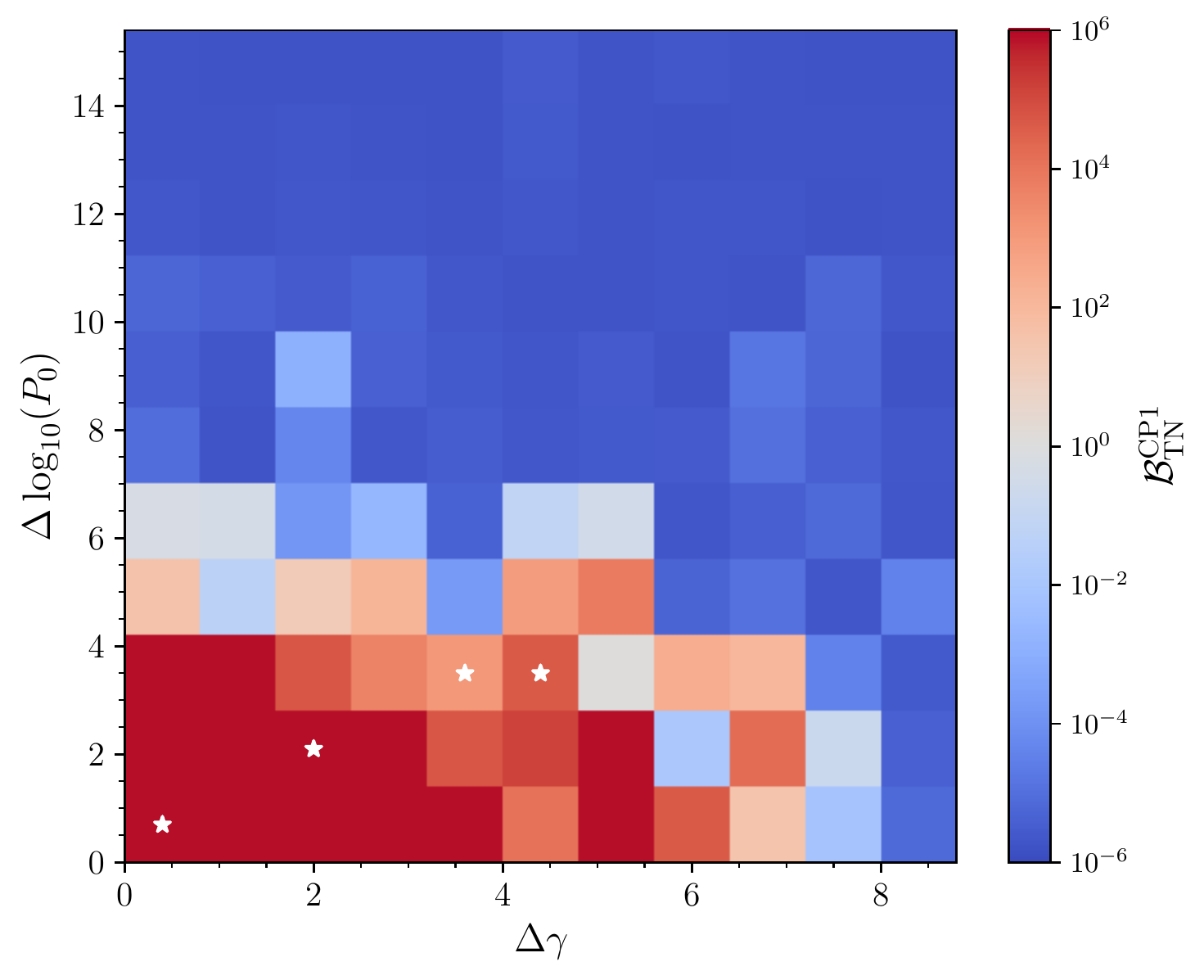}
    \includegraphics[width=\linewidth]{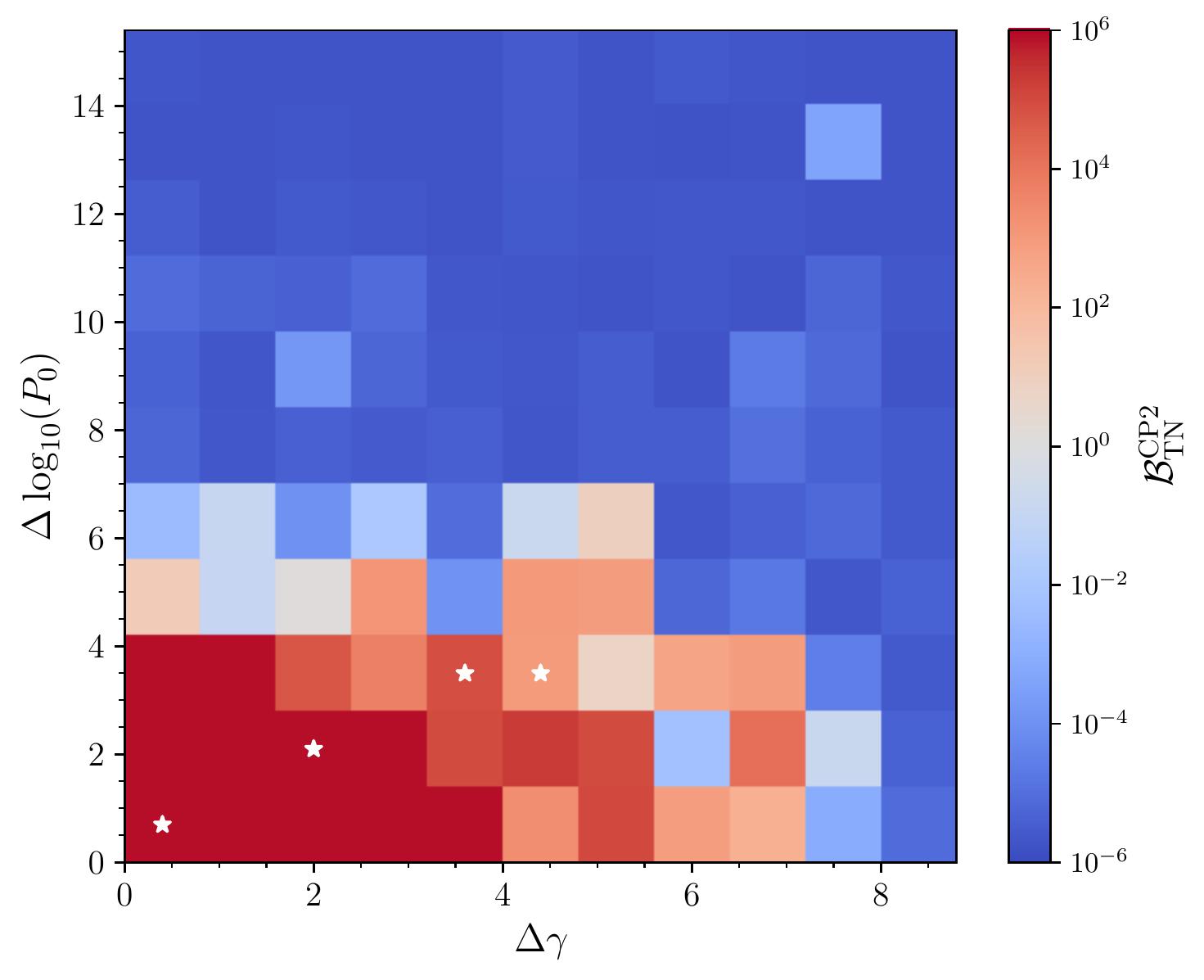}
    \caption{realization-averaged Bayes factors for model $\mathrm{CP1}$ (top) and $\mathrm{CP2}$ (bottom) over $\mathrm{TN}$, as a function of $\Delta \log_{10} P_0$ and $\Delta \gamma$. Both $\mathrm{CP1}$ and $\mathrm{CP2}$ are heavily favoured across a wide range of variations in input timing noise parameters. White stars indicate the samples of the $(\Delta \log_{10} P_0, \Delta \gamma)$ parameter space plotted in Figure \ref{fig:noise_PSDs}.}
    \label{fig:bf_matrix}
\end{figure}

\begin{figure}
    \centering
    \includegraphics[width=\linewidth]{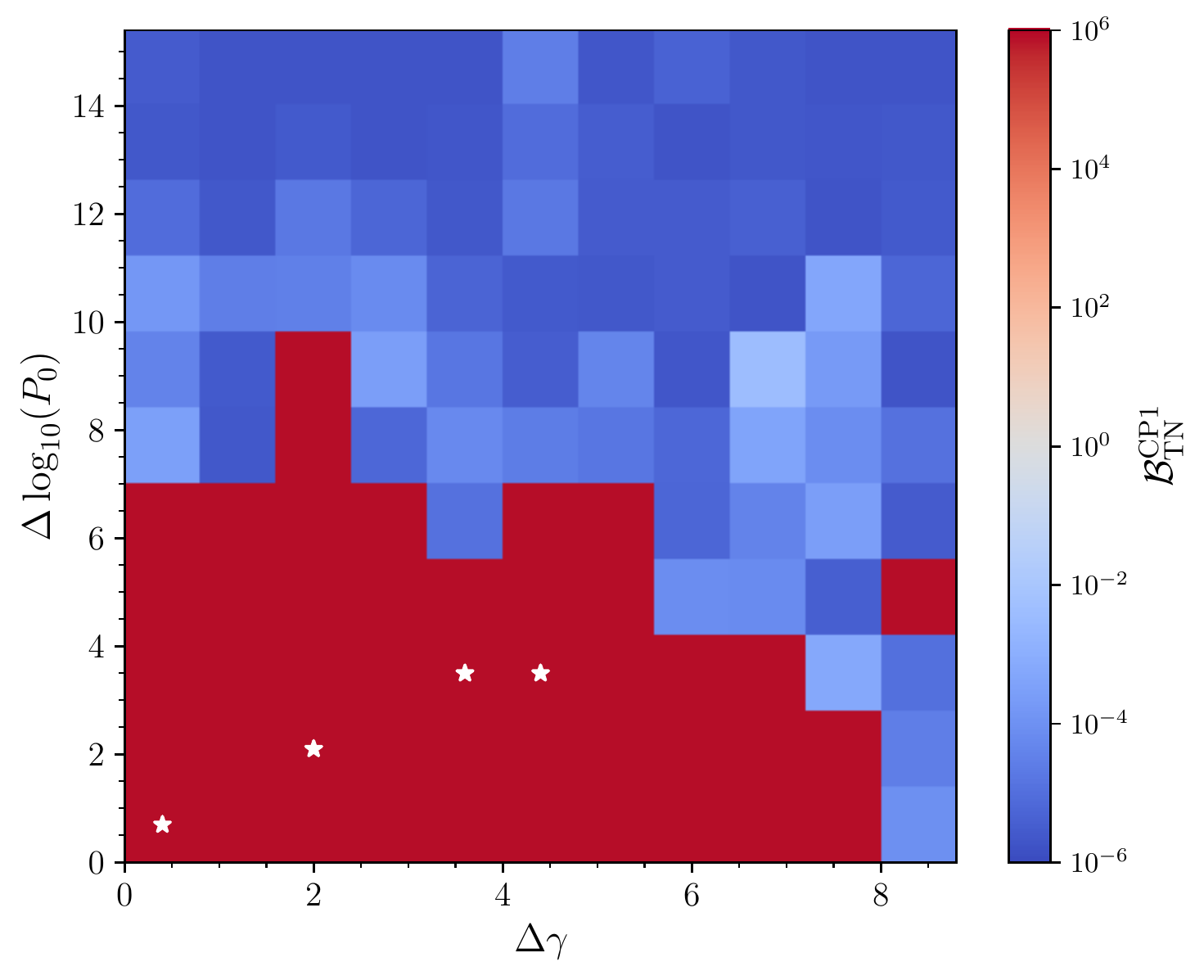}
    \includegraphics[width=\linewidth]{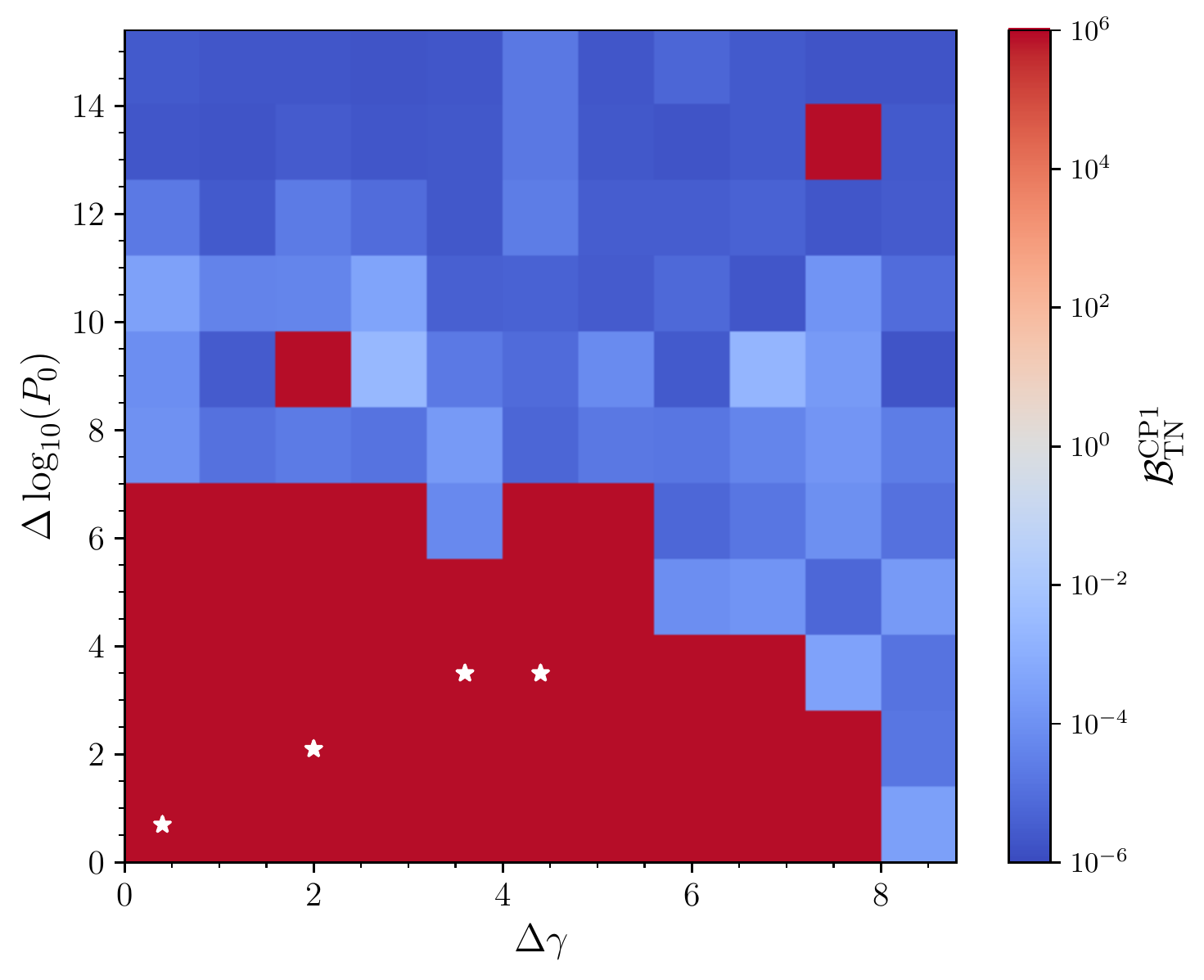}
    \caption{Same as Figure \ref{fig:bf_matrix}, but for the maximum Bayes factors across all noise realizations.}
    \label{fig:max_bf_matrix}
\end{figure}

We used an established Bayesian inference procedure \citep[see ][]{2022MNRAS.510.4873A,2021MNRAS.508.4970C,2021ApJ...917L..19G, 2020ApJ...905L..34A} for our analysis. To summarize, we use the multivariate Gaussian likelihood to model the data \citep{2017PhRvD..95d2002T,2016ApJ...821...13A}. Next, we construct the so-called design matrix with \textsc{libstempo}~\citep{2020ascl.soft02017V} and \textsc{tempo2}~\citep{2006MNRAS.372.1549E} to marginalize the likelihood over the terms in the deterministic pulsar timing model. 
We used \textsc{enterprise} \citep[][]{2019ascl.soft12015E} to perform Bayesian model selection and parameter estimation using \textsc{ptmcmcsampler} \citep{2019ascl.soft12017E}. We used the hybrid Bayesian-frequentist optimal statistic \citep[][see Section \ref{sec:os}]{2009PhRvD..79h4030A,2013ApJ...762...94D,2015PhRvD..91d4048C} to evaluate evidence for spatial correlations in the simulated datasets.

We employed three models to evaluate the simulated datasets in our analysis:
\begin{itemize}
    \item $\mathrm{TN}$: Independent timing noise for each pulsar, parametrized by a red power-law spectrum as in Eq. \ref{eq:rednoise_spec} (this is generally the correct description, except in simulations with $(\Delta \log_{10} P_0, \Delta \gamma ) = (0,0)$, which can be described as a true spatially-uncorrelated common-spectrum process).
    \item $\mathrm{CP1}$: Independent red-spectrum timing noise (as with model $\mathrm{TN}$) for each pulsar, but with the addition of a common-spectrum noise process with varying spectral index $\gamma_{\text{CP}}$ and amplitude $A_\mathrm{CP}$.
    \item $\mathrm{CP2}$:\footnote{When referring to either model CP1 and CP2 generally, we use CP.} Same as model $\mathrm{CP1}$, but with the CRN spectral index $\gamma$ held fixed to the fiducial value of $13/3$ expected for a classical GWB.
\end{itemize}

The red noise terms were evaluated using a Fourier series with a fundamental frequency corresponding to the inverse of the dataset observing time-span $1/T_\mathrm{obs}$. We held white noise hyper-parameters (known in the pulsar timing community as $\mathrm{EFAC}$ and $\mathrm{EQUAD}$) \revision{fixed at 1 and 0 respectively}, as our simulations did not incorporate any deviations of the white noise characteristics from the injected values.

In our standard Bayesian posterior sampling and model selection runs, we used uniform priors on the timing noise and CRN spectral indices ($\gamma \in U[0,10]$), and log-uniform priors on the timing noise and CRN amplitudes ($\log A \in U[-20, -6])$.

\subsection{Model selection analysis}

We performed model selection for models $\mathrm{CP1}$ and $\mathrm{CP2}$ over $\mathrm{TN}$ using the product-space method in a ``hypermodel'' framework \citep{carlinchibb,2016MNRAS.455.2461H,2020PhRvD.102h4039T}.
To provide adequate dynamic range for measuring very large or very small Bayes factors, we sampled with $1\times 10^{6}$ iterations. Because of computational costs, we only analysed ten noise realizations for each cell across the $(\Delta \log_{10} P_0, \Delta \gamma)$ parameter space, meaning that we only processed 1210 out of 12100 simulated datasets for the model selection analysis. After performing the model selection procedure on each noise realization, we computed the realization-averaged Bayes factor in log-space.

The results from this search are shown in Figures \ref{fig:bf_matrix} and \ref{fig:max_bf_matrix}. On average, we find strong support for model $\mathrm{CP1}$ and $\mathrm{CP2}$ over $\mathrm{TN}$, for a region of parameter space spanning up to $\Delta \gamma \sim 4$ and $\Delta \log_{10} P_0 \sim 4$ ($\log_{10} \mathcal{B}^{\text{CP}}_{\text{TN}} \gtrsim 5.8$), and more moderate support for values of $\Delta \log_{10} P_0$ up to 6, and $\Delta \gamma$ up to 7. That is, CP models remain the preferred model on average over TN under $\sim $~six orders-of-magnitude variations in timing noise amplitude, and variations in the timing noise spectral index by a range of $\sim 7$. 
In Figure \ref{fig:max_bf_matrix}, we show the maximum Bayes factors across all realizations. We find that the Bayes factors are limited by the number of posterior samples over the entire region in Figure \ref{fig:bf_matrix} where the CP models are not strongly disfavoured. This indicates that even if there is moderate support on average for a CRN process for a given choice of $(\Delta \log_{10} P_0, \Delta \gamma)$, strong support for CP models is found in at least one out of ten realizations.

Our simulations demonstrate that model comparison of CP1 or CP2 against TN alone is insufficient to claim detection of CRN.
If the inference of a CRN was to be useful as preliminary evidence for a GWB, then the CRN models (CP1 and CP2) should only be favoured in our simulations when $\Delta \log_{10} P_0$ and $\Delta \gamma$ are close to 0. Furthermore, the support should quickly decline as $\Delta \log_{10} P_0$ and $\Delta \gamma$ increase. Instead, we see a more gradual decline on average as the span of timing noise parameters increases.

\subsubsection{The distribution of Bayes factors}
\revision{When comparing appropriately-specified cosmological models, Bayes factors can be expected to exhibit scatter of about an order of magnitude due to cosmic variance \citep[e.g.][]{2021A&A...647L...5J}. In principle, this scatter may be used to set appropriate (and more conservative) decision thresholds in Bayesian model comparison. Similar boot-strapping approaches have already been developed for frequentist detection statistics for spatial correlations in pulsar timing array analysis \citep{2017PhRvD..95d2002T}. We investigated the underlying Bayes factor distributions to determine whether it is possible to calibrate Bayes factors for the CRN under current procedures. To improve our sample statistics, we group cells in the $\Delta \log_{10} P_0, \Delta \gamma$ parameter space by average Bayes factor values.} 

\begin{figure}
    \centering
    \includegraphics[width=\linewidth]{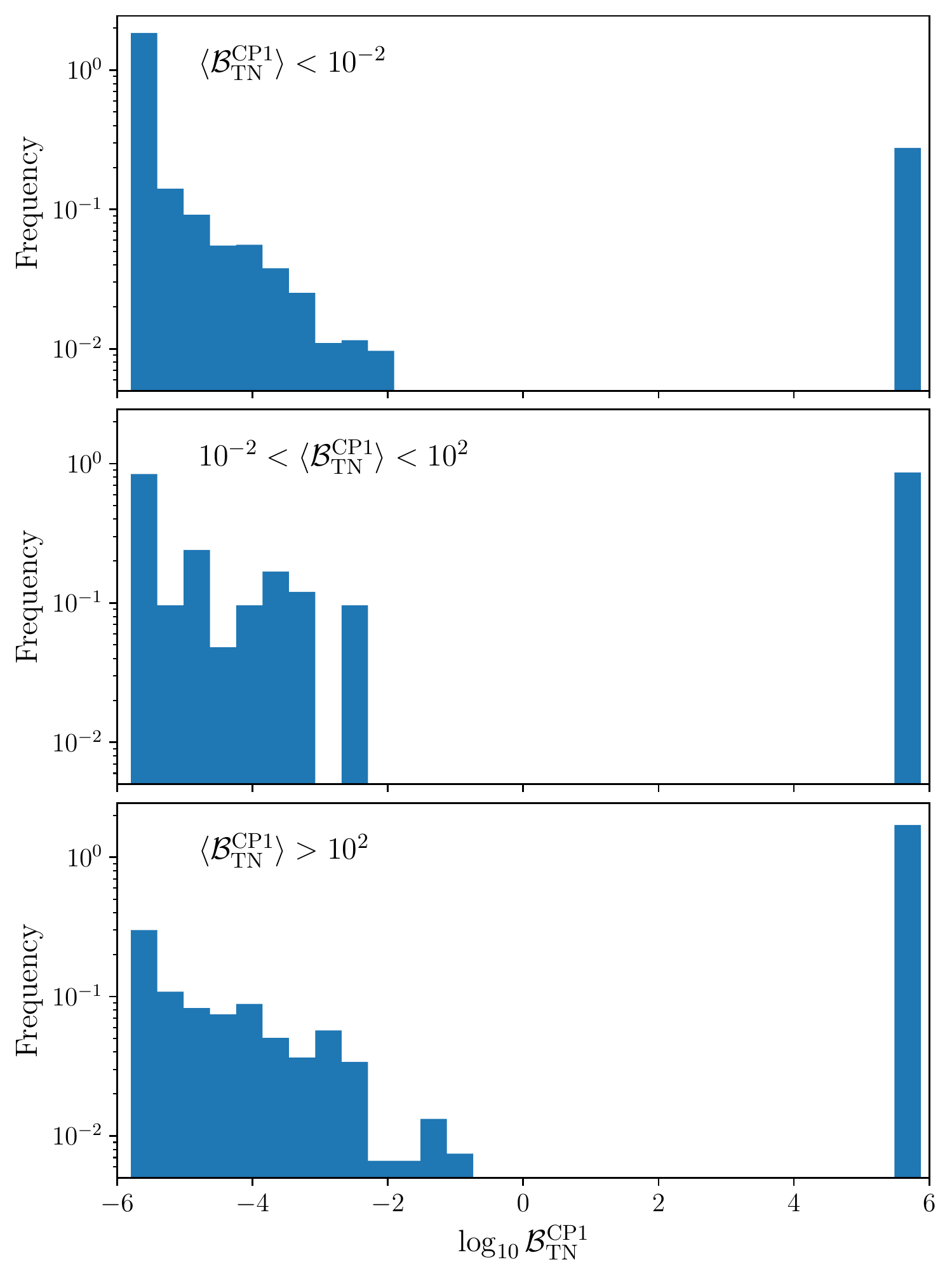}
    \caption{\revision{Relative distributions of Bayes factors for CP1 over TN, grouped by realisations with $\mathcal{B}^{\mathrm{CP1}}_{\mathrm{TN}} < 10^{-2}$ (top), $10^{-2}< \mathcal{B}^{\mathrm{CP1}}_{\mathrm{TN}} < 10^{2}$ (middle), $\mathcal{B}^{\mathrm{CP1}}_{\mathrm{TN}} > 10^{2}$ (bottom)}.}
    \label{fig:bf_distributions}
\end{figure}

\revision{We show Bayes factor distributions in Figure \ref{fig:bf_distributions}, for $\Delta \log_{10} P_0, \Delta \gamma$ cells where $\mathcal{B}^{\text{CP1}}_{\text{TN}}< 10^{-2}$ (top), $10^{-2} <\mathcal{B}^{\text{CP1}}_{\text{TN}}< 10^{2}$ (middle), $ \mathcal{B}^{\text{CP1}}_{\text{TN}} > 10^{2}$ (bottom). These groupings represent cells where CP is, on average, strongly disfavored, weakly disfavored/favored, and strongly favored (respectively). While a minority of Bayes Factors have values representing moderate to strong evidence against CP1, most Bayes Factors in the sample are peaked at the boundary values close to $10^{\pm 6}$, which are set by the number of our posterior samples in the hypermodel framework. Furthermore, there are very few Bayes factors within intermediate values, with only a small tail weighted toward $\mathcal{B}_\text{TN}^{\text{CP1}} < 1$. These features suggest that many more posterior samples are required to resolve the true underlying Bayes factor distributions. More importantly, these distributions highlight the improper performance of CRN inference under current models, priors, and model selection procedures. While further investigations of these underlying Bayes factor distributions may be a topic of interest for future work, we suggest that improvements in the underlying inference procedure \citep[e.g.][]{2022ApJ...932L..22G} is a more appropriate route toward robust CRN inference in future.}

\subsubsection{The effect of prior volumes}

We consider the possibility that the spurious support for CP models arises from the choice of priors on timing noise parameters. This is motivated by the fact that when a CRN is well-constrained in our simulations, the estimated timing noise amplitudes for most pulsars tend to drop to very small values, and the spectral indices become unconstrained. In our standard analyses, and the analysis presented in \citet{2021ApJ...917L..19G}, we use wide priors on timing noise parameters. This is reflective of our lack of \textit{a-priori} knowledge of the timing noise properties of the pulsars, particularly since the spectral parameters for timing noise and the GWB are highly covariant.

If the priors on timing noise parameters are chosen such that their range more closely reflects the true range of injected values, the support for CP models may diminish. 
To test this, we selected simulated datasets with $(\Delta \log_{10} P_0, \Delta \gamma) = (1.4, 0.8)$ and $(2.8, 1.6)$. As before, we performed our model selection analysis on 10 realizations, but this time with a gradually decreasing prior width on $\log A$ and $\gamma$, until the priors approached the delta function at the median timing noise parameters. 

\begin{figure}
    \centering
    \includegraphics[width=\linewidth]{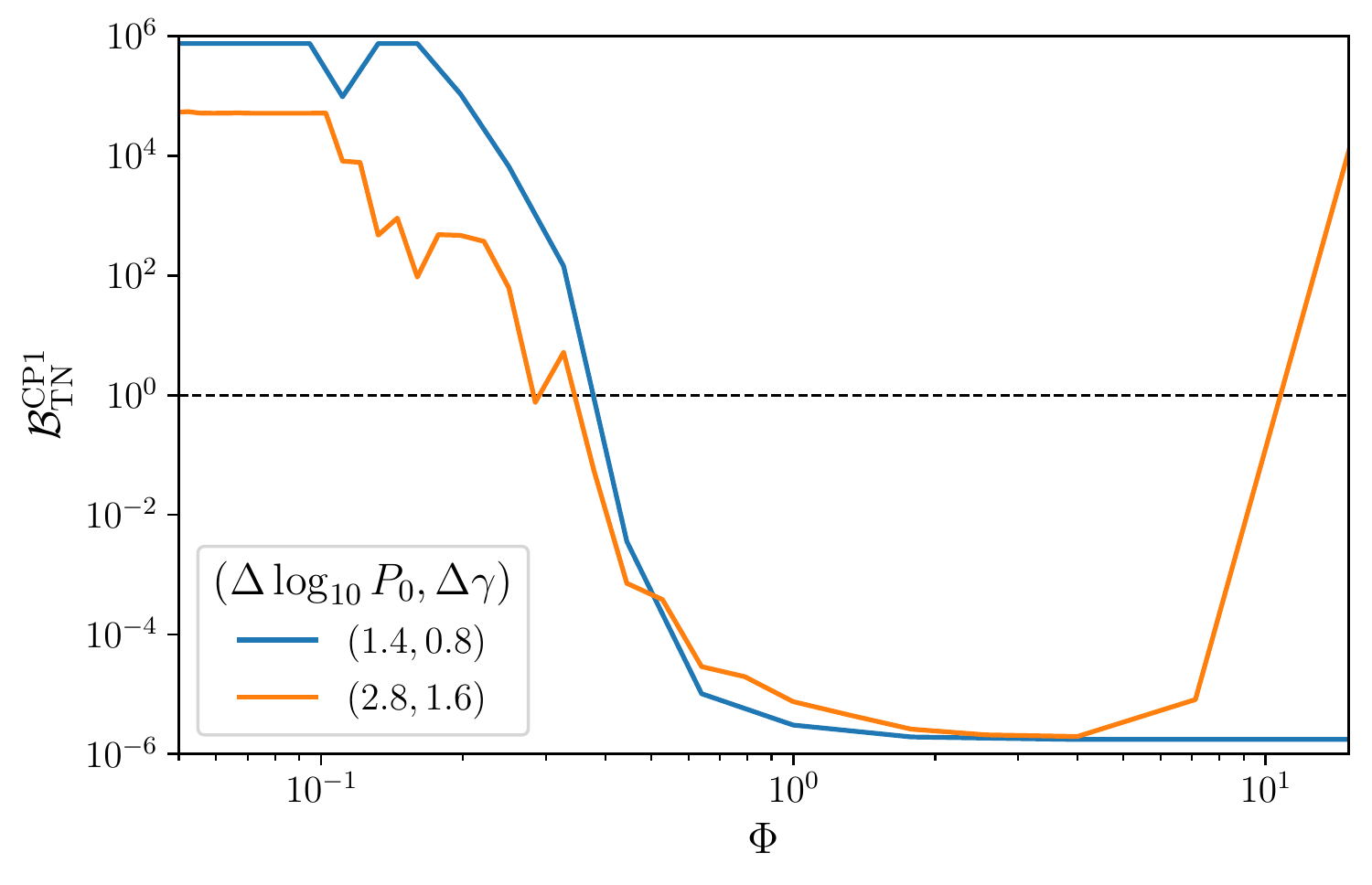}
    \caption{realization-averaged Bayes factors for model CP1 over TN as a function of the ratio between the injected and prior timing noise parameter space volumes, $\Phi$, for $(\Delta \log_{10} P_0, \Delta \gamma) = (1.4, 0.8)$ (blue) and $(2.8, 1.4)$ (orange). The black dashed line indicates a Bayes Factor of 1.}
    \label{fig:prior_bf_plot}
\end{figure}

The results are shown in Figure \ref{fig:prior_bf_plot}, where we plot realization-averaged Bayes factors for CP2 over TN, as a function of $\Phi~=~(\Delta \log_{10} A \Delta \gamma)_\text{inj} / (\Delta \log_{10} A \Delta \gamma )_\text{prior}$ -- the ratio between the volumes of simulated and prior timing noise parameter distributions. Heuristically, when the ratio $\Phi$ is small, the priors span a wider range than the range of injected timing noise parameters, and vice versa for large values of $\Phi$. Figure \ref{fig:prior_bf_plot} shows that as the ratio approaches unity, the support for the CRN quickly diminishes.

For simulations with a wider range of timing noise prescriptions, such as $(\Delta \log_{10} A, \Delta \gamma ) = (2.8, 1.6)$ shown in Figure \ref{fig:prior_bf_plot}, Bayes factors begin to increase again at large values of $\Phi$. This is likely because the timing noise priors are too restrictive given the variance of timing noise prescriptions in the simulated datasets, causing the CP model to be favoured over TN once again (even though neither model represents a good description of the data with the choice of priors). Overall, these effects highlight the sensitivity of CRN inference to the choice of priors on timing noise parameters.

It is not possible to accurately bound the priors on timing noise parameters to match the true distributions over all pulsars in real datasets -- we do not know the true underlying distribution of pulsar timing noise parameters \textit{a-priori}, particularly for low-amplitude timing noise. As mentioned above, red noise in pulsar timing residuals could be ascribed to either timing noise, or a GWB signal, or a combination of these. The results presented above are simply an exercise in demonstrating the sensitivity of current methodologies to the choice of priors.

\subsection{Common-spectrum process and timing noise characteristics}
\begin{figure}
    \centering
    \includegraphics[width=\linewidth]{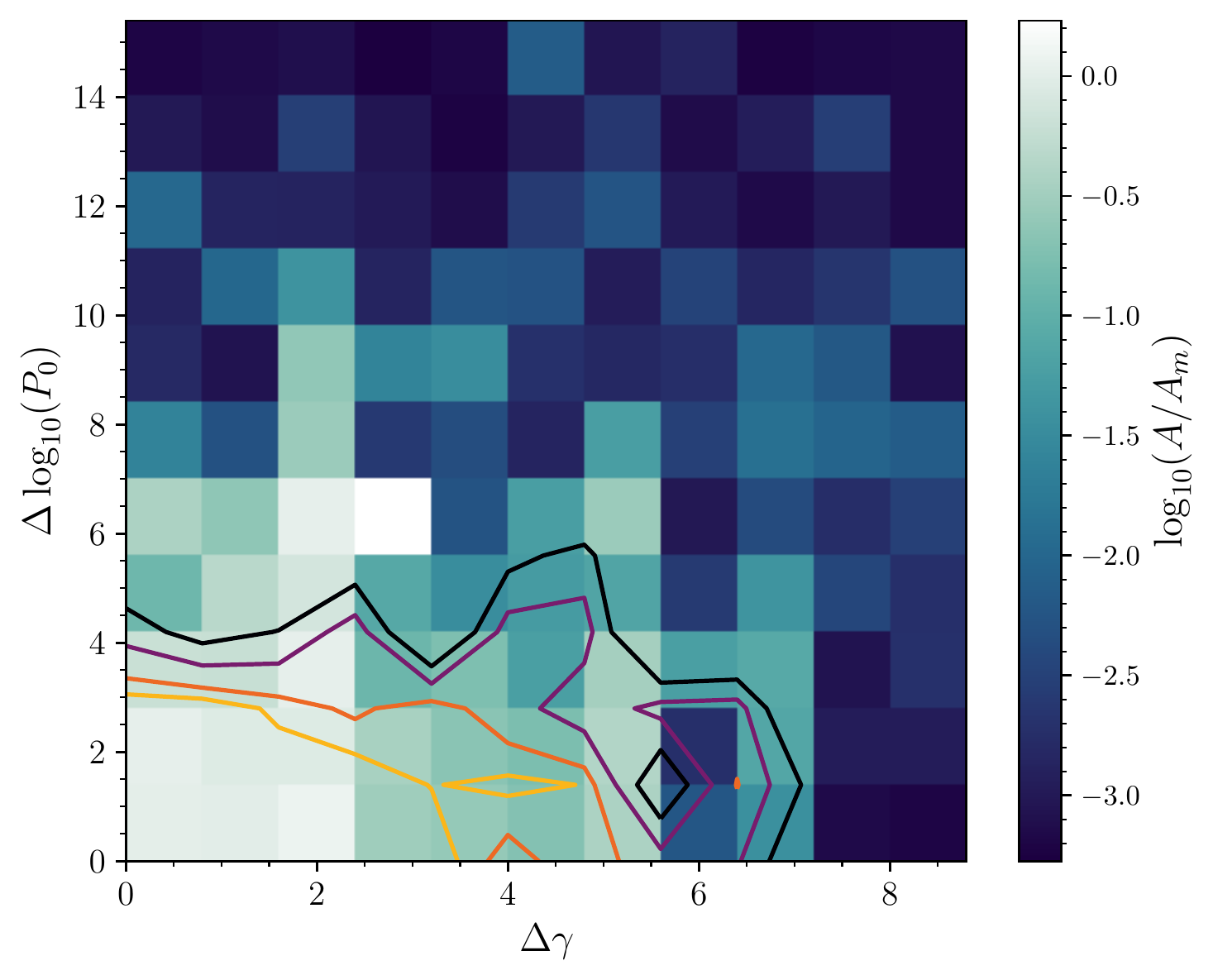}
    \includegraphics[width=\linewidth]{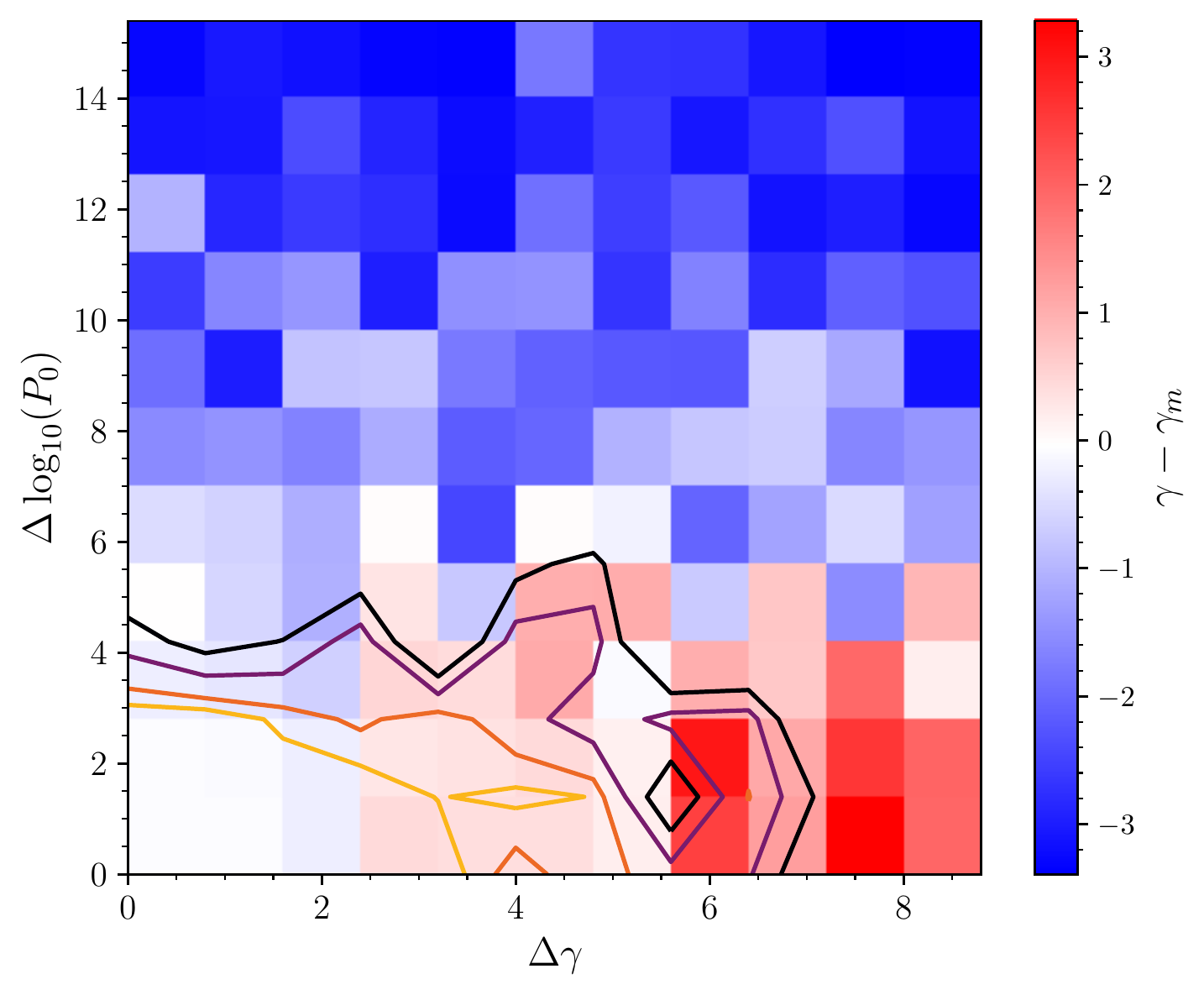}
    \caption{Difference between the inferred CRN amplitude (top) and spectral index (bottom) and the median values of injected timing noise parameters. Contours showing Bayes factors for CP1 over TN are over-plotted, with levels at $\log_{10} \mathcal{B}^{\text{CP1}}_{\text{TN}} = 6, 5, 4, 3$ (in order of light to dark).}
    \label{fig:dA_dgam}.
\end{figure}

%In \citet{2021ApJ...917L..19G}, we identified that the CRN can closely match the .
To better understand the origin of spurious CRN detections, we now consider the relationship between the CRN inferred in our standard analysis and the injected timing noise. We performed Bayesian parameter estimation for model CP1 across the timing noise parameter space. In Figure \ref{fig:dA_dgam} we show the difference between the posterior-median CRN parameters and the central values of the injected timing noise parameters, $\log_{10} A_m = -14.46$, and $\gamma_m = 4$. In the region of parameter space where CP1 is preferred, the recovered CRN is close to the central timing noise parameters, indicating that the inferred CRN is consistent with the ensemble average of independent pulsar timing noise terms. 

This is perhaps not surprising, considering the heuristic description of the CRN as a spectral process present among all pulsars, and suggests that timing noise can bias estimates for a CRN when it is lower in amplitude than typical timing noise terms. Indeed, in \citet{2021ApJ...917L..19G}, we showed that a CRN signal need not be present in all pulsars to be inferred -- some pulsars have a substantially lower red noise level than the majority of pulsars in both real and simulated PTA datasets, but this has little impact on the detection of an apparent CRN signal.

Some trends in the characteristics of the inferred CRN are apparent. In Figure \ref{fig:dA_dgam} bottom, a systematic trend of steeper spectral indices, and lower amplitudes for the inferred common spectrum process is evident when $\Delta \gamma$ is large, $\Delta \log_{10} P_0$ is small. Similarly, when $\Delta \log_{10} P_0$ is large and $\Delta \gamma$ is small, the inferred CRN spectrum tends to be slightly shallower, and slightly lower in amplitude.

\subsection{Optimal statistic analysis for spatial correlations}

\subsubsection{The optimal statistic}
\label{sec:os}
It is important to consider whether spurious detections of a CRN may also result in spurious detections of spatial correlations, and if so, how often. To do this, we employed the optimal statistic, $\hat{A}^{2}$  \citep{2009PhRvD..79h4030A,2013ApJ...762...94D,2015PhRvD..91d4048C,2018PhRvD..98d4003V}, which is a frequentist estimator of the amplitude of spatially-correlated noise processes. It is constructed as the weighted sum of inter-pulsar spatial correlations accounting for pulsar-specific and inter-pulsar noise covariances, and is given by \citep{2015PhRvD..91d4048C}
\begin{equation}
    \hat{A}^{2} = \frac{\sum_{ab} \delta \mathbf{t}_{a}^{T} C_a^{-1} \tilde{S}_{ab} C_b^{-1} \delta \mathbf{t}^{b}}
    {\sum_{ab} \mathrm{tr}\left(C_a^{-1} \tilde{S}_{ab} C_b^{-1} \tilde{S}_{ba}\right)}\,\text{,}
\end{equation}
where $\delta \mathbf{t}_{a}$ is the vector of timing residuals for pulsar $a$, $C_a = \langle \delta \mathbf{t}_{a} \delta \mathbf{t}_{a}^{T}\rangle$ is the autocovariance matrix, $\tilde{S}_{ab} = S_{ab} A_\text{GWB}^{-2}$ is the GWB amplitude-normalized cross correlation matrix, with $S_{ab} = \langle \delta \mathbf{t}_{a} \delta \mathbf{t}_{b}^{T}\rangle |_{a \neq b}$. 
The optimal statistic signal-to-noise (S/N) ratio gives a measure of the significance for $A_\text{GWB} \neq 0$, and is given by
\begin{equation}
    \rho = \frac{\sum_{ab} \delta \mathbf{t}_{a}^{T} C_a^{-1} \tilde{S}_{ab} C_b^{-1} \delta \mathbf{t}^{b}}
           {\left[\sum_{ab} \mathrm{tr}\left(C_a^{-1} \tilde{S}_{ab} C_b^{-1} \tilde{S}_{ba}\right)\right]^{1/2}}\,\text{.}
\end{equation} 

In the standard approach, the pulsar-intrinsic red noise terms are first jointly sampled with the CRN terms using Bayesian parameter estimation, and held fixed at the maximum-likelihood values when computing the optimal cross-correlation statistic \citep{2015PhRvD..91d4048C}. However, this method does not fully account for the degeneracy between pulsar-intrinsic red noise and the red noise induced by the GWB, resulting in biased estimates of the GWB amplitude \citep{2018PhRvD..98d4003V}. To address this, \citet{2018PhRvD..98d4003V} developed the noise-marginalized optimal statistic, which estimates the cross-correlation amplitude using posterior samples from the pulsar-intrinsic noise terms from the joint pulsar-intrinsic and CRN parameter estimations (e.g., parameter estimation of model CP2). The optimal statistic has been deployed as a complement to fully Bayesian characterisation of inter-pulsar spatial correlations in PTA datasets \citep{2018ApJ...859...47A, 2020ApJ...905L..34A, 2022MNRAS.510.4873A}. The results from optimal statistic analysis of spatial correlations are broadly consistent with the fully Bayesian measurements of spatial correlations, but there are some key differences; namely, the presence of a monopole-correlated signal is marginally supported in optimal statistic analyses of recent PTA datasets \citep{2022MNRAS.510.4873A,2020ApJ...905L..34A}, but is not supported by Bayesian analyses.%This will be the subject of an upcoming work.

\subsubsection{Optimal statistic false detection analysis}
To assess the robustness of spatial correlation inference in the presence of a wide range of timing noise characteristics, we computed the optimal statistic for HD, dipole, and monopolar correlations on our simulated datasets containing only pulsar timing noise. We used posterior samples from Bayesian parameter estimation runs of model CP2 over our simulated datasets to calculate both the standard (maximum-likelihood) and noise-marginalized optimal statistics. In both cases, we also calculated the optimal statistic S/N $\rho$ and the inter-pulsar covariance measured in angular bins to investigate the significance of any false detections.

\begin{figure}
    \centering
    \includegraphics[width=0.97\linewidth]{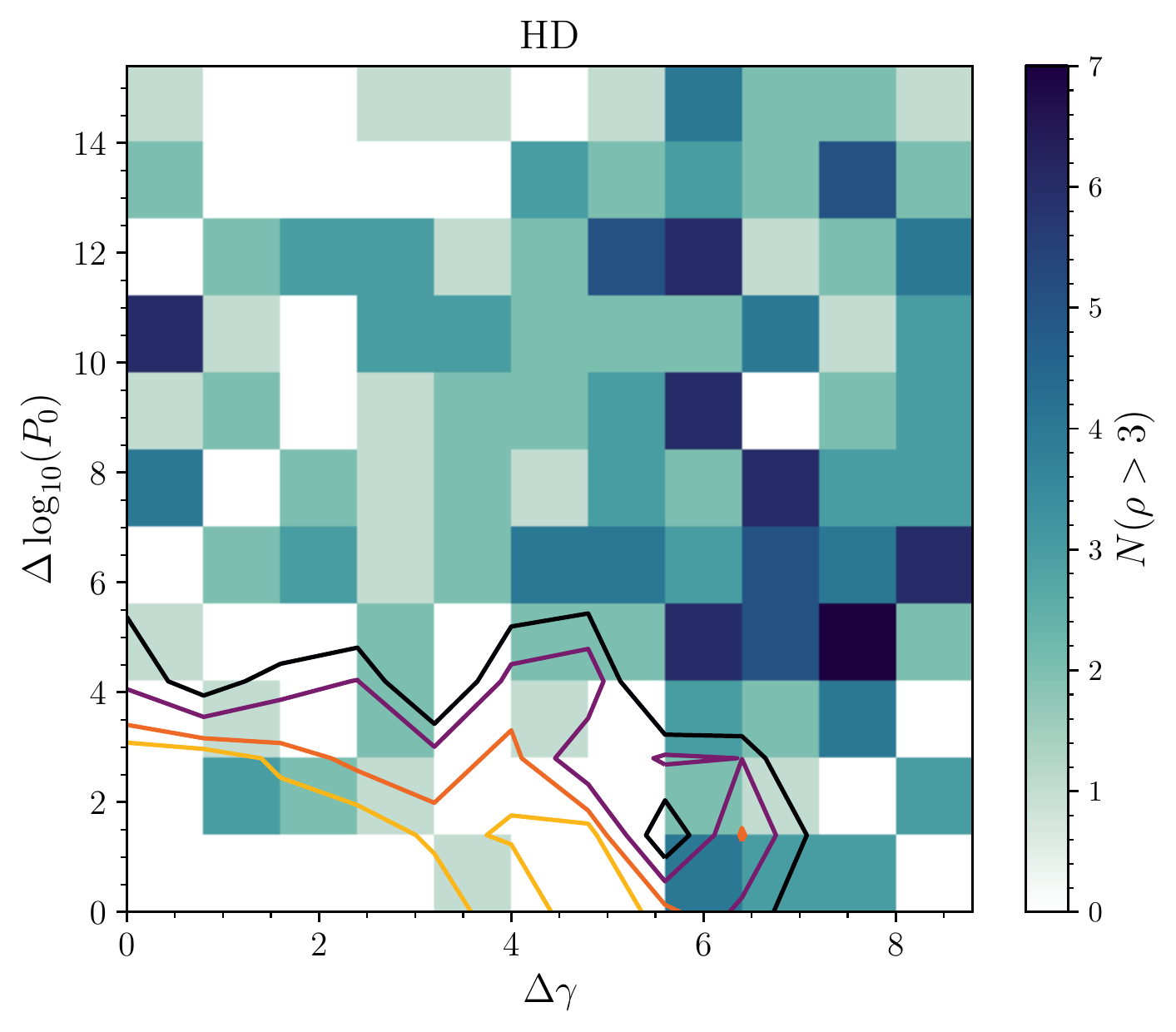}
    \includegraphics[width=0.97\linewidth]{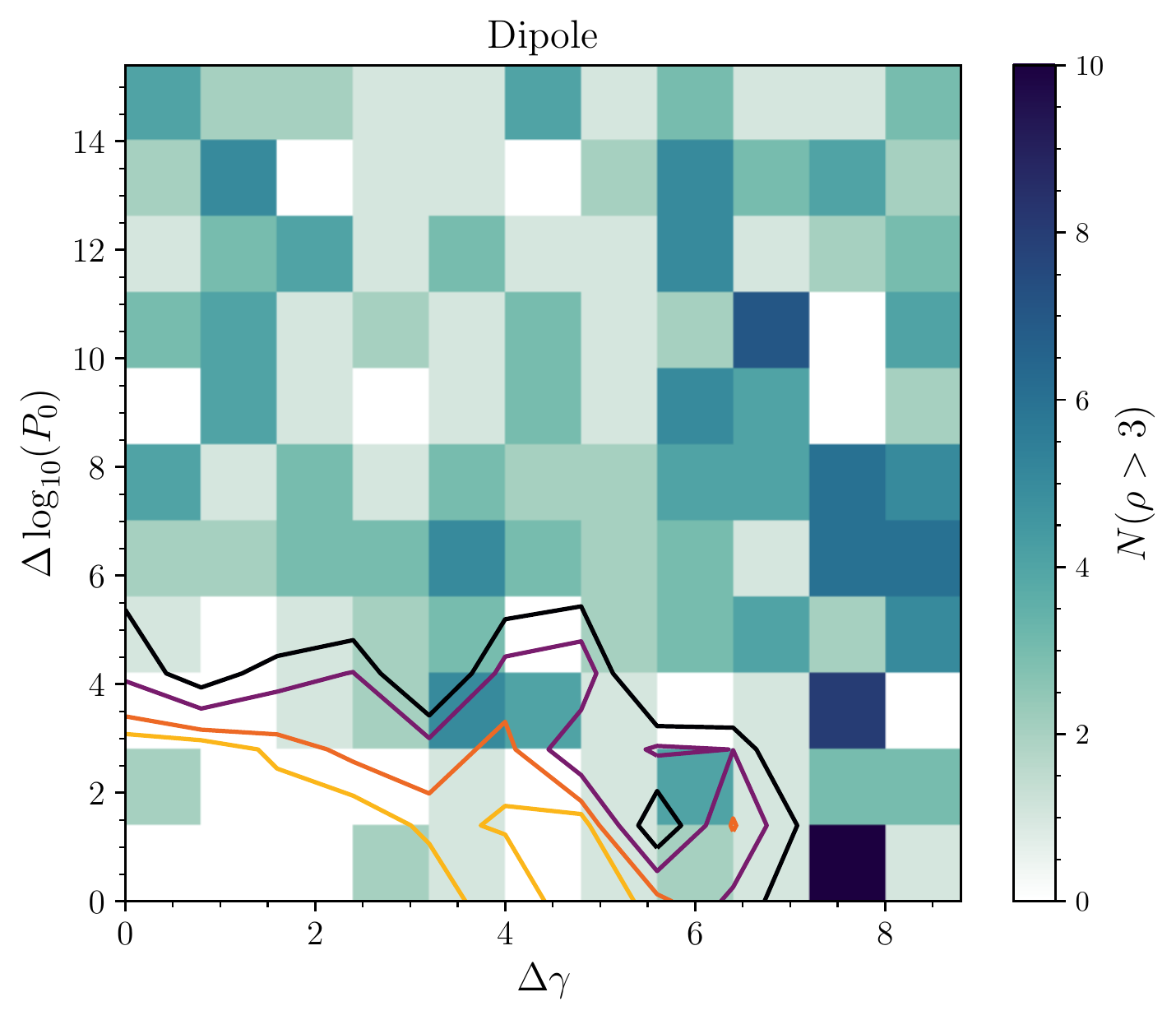}
    \includegraphics[width=0.97\linewidth]{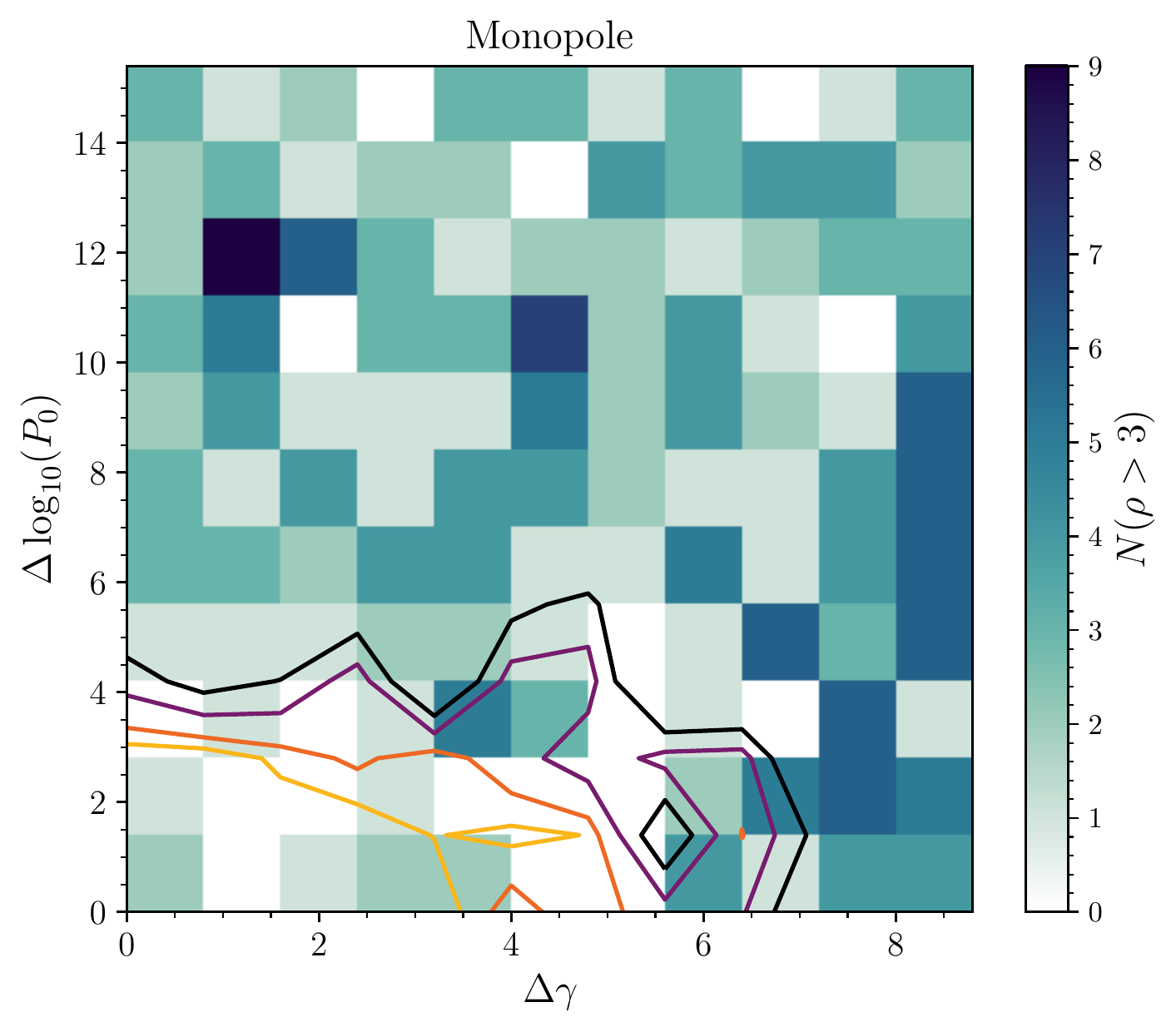}
    \caption{Percentage of realizations with a maximum-likelihood optimal statistic S/N greater than 3, for HD (top), dipolar (middle), and monopolar (bottom) spatial correlations, as a function of the variation in input timing noise parameters, $\Delta \log_{10} P_0$ and $\Delta \gamma$. Contours are as in Figure \ref{fig:dA_dgam}}
    \label{fig:os_sn3}
\end{figure}

\begin{figure}
    \centering
    \includegraphics[width=0.98\linewidth]{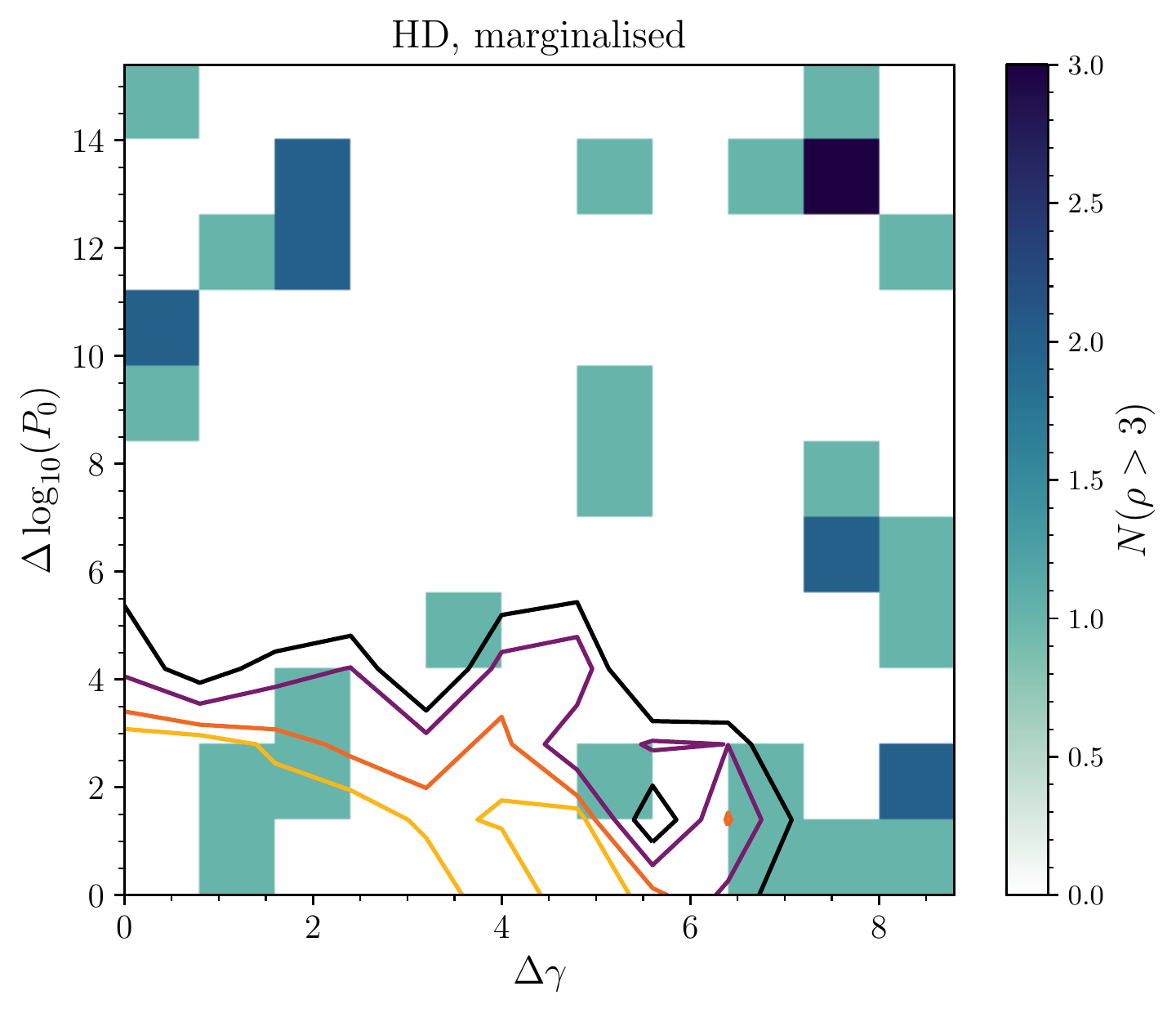}
    \includegraphics[width=0.98\linewidth]{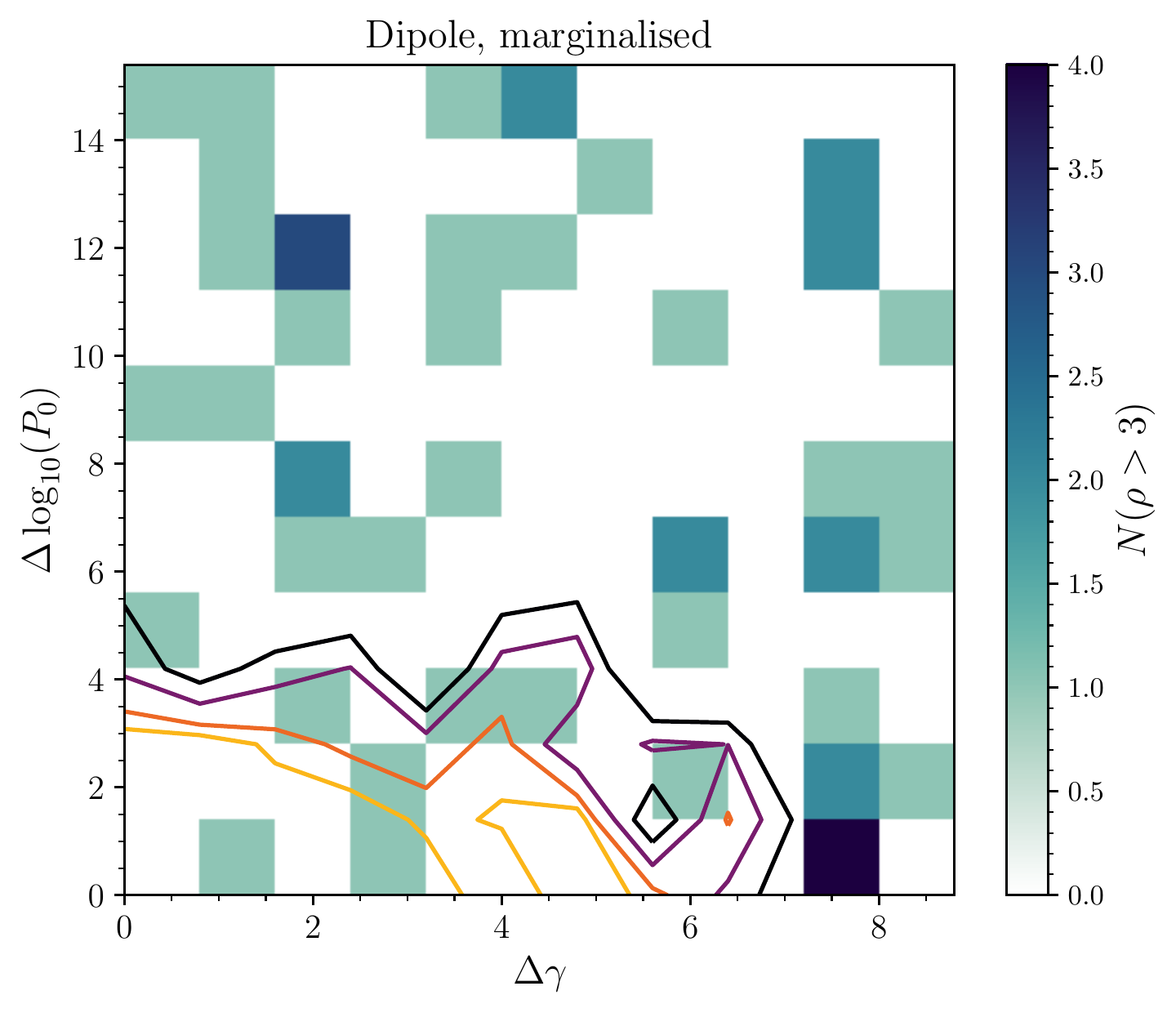}
    \includegraphics[width=0.98\linewidth]{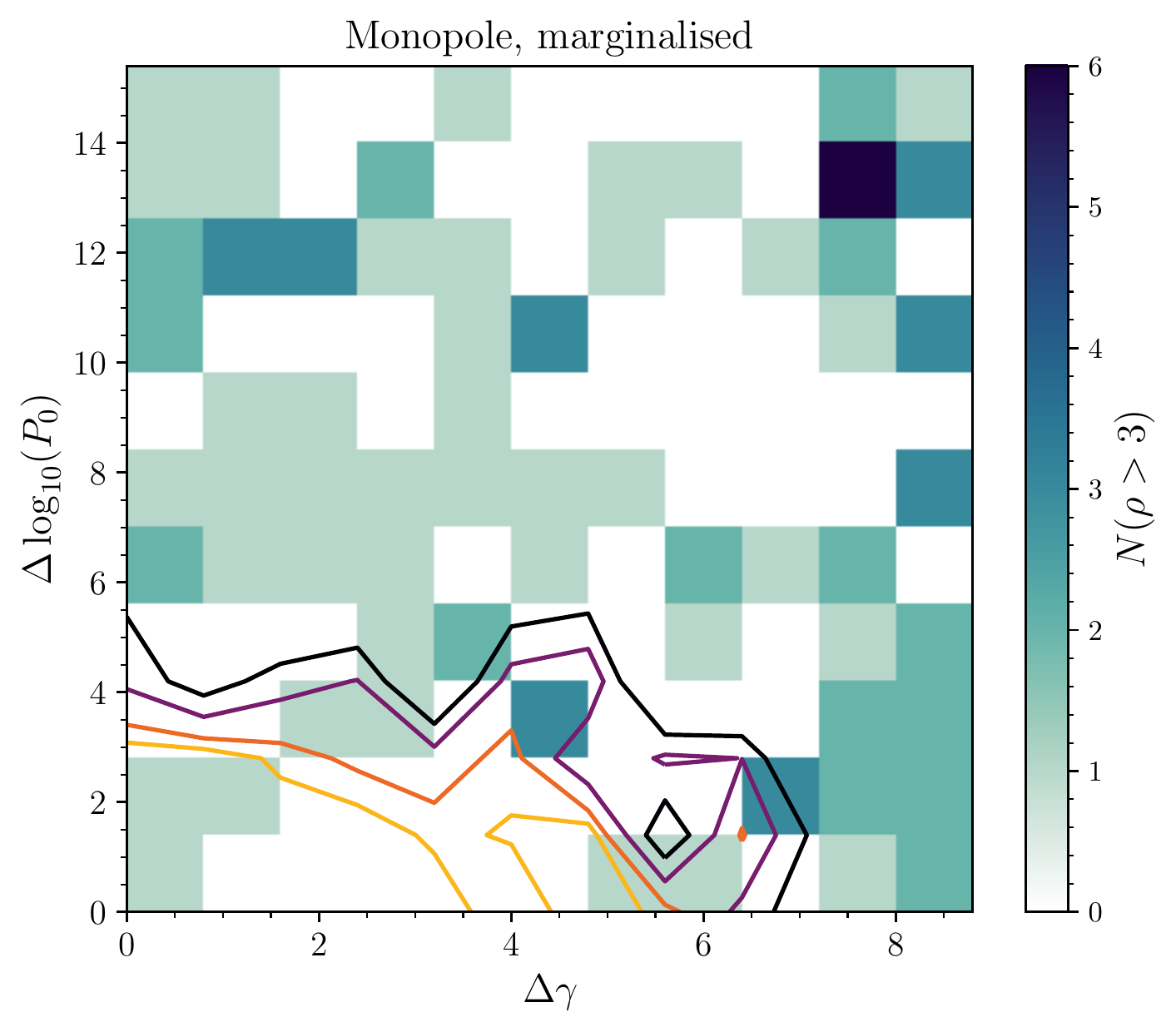}
    \caption{Same as Figure \ref{fig:os_sn3}, but for the noise-marginalized optimal statistic.}
    \label{fig:os_sn3_marg}
\end{figure}

\begin{figure}
    \centering
    \includegraphics[width=\linewidth]{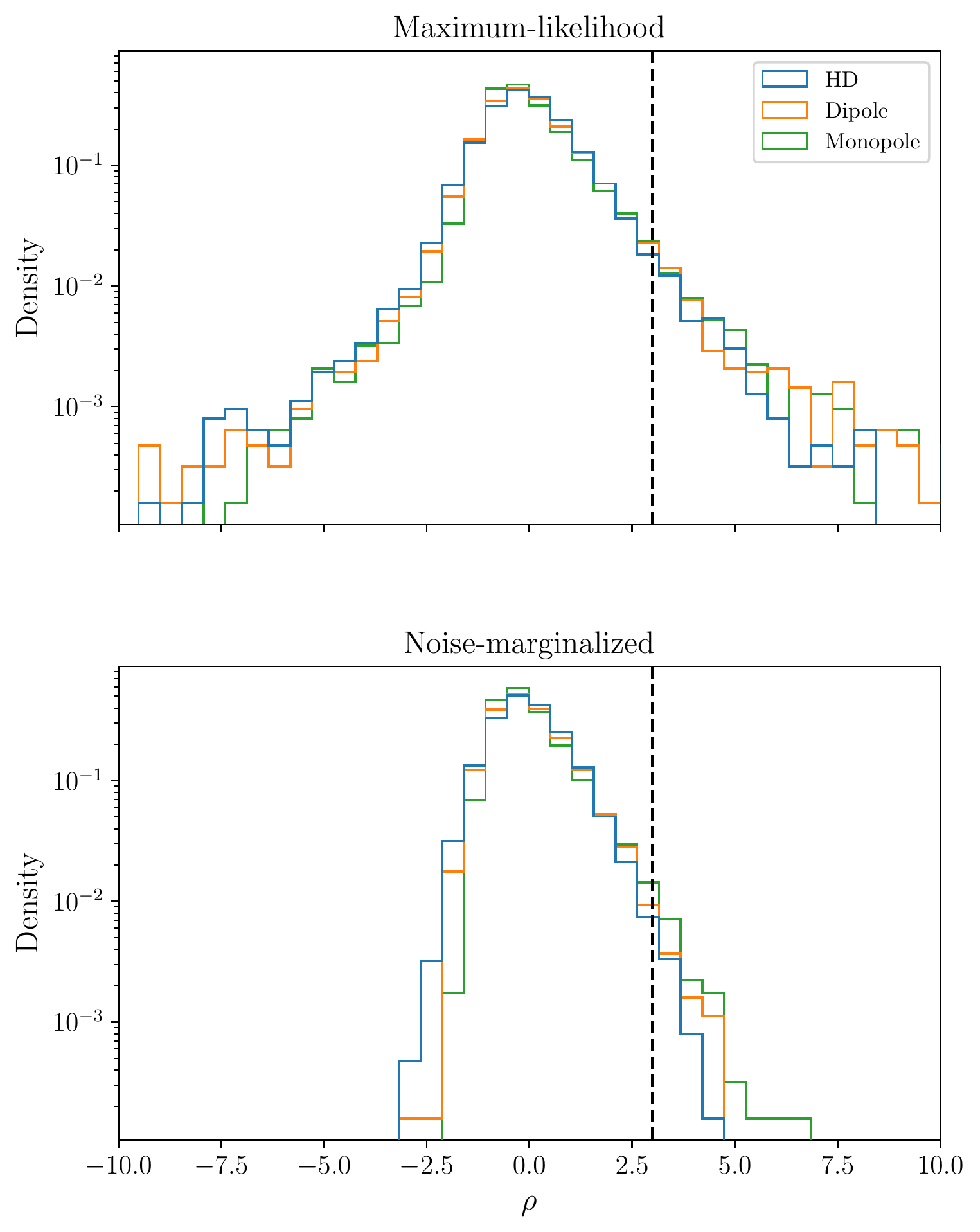}
    \caption{Distribution of S/N ratios $\rho$ for HD, dipole, and monopole correlations, for the maximum-likelihood (top) and mean noise-marginalized (bottom) optimal statistic, over the full $(\Delta \log_{10} P_0$, $\Delta \gamma)$ parameter space. The black dashed line indicates a S/N threshold of 3.}
    \label{fig:os_snr_hist}
\end{figure}

In Figures \ref{fig:os_sn3} and \ref{fig:os_sn3_marg} we show the number of realizations (out of 100) with $\rho > 3$ for the maximum-likelihood and noise-marginalized optimal statistic (respectively), as a proxy for the number of false detections of spatial correlations. We also show the overall fraction of realizations with $\rho > 3$ over all values of $\Delta \log_{10} P_0$ and $\Delta \gamma$ in Table \ref{tab:os_ndetections}. Figure \ref{fig:os_snr_hist} shows the overall distribution of $\rho$ for the maximum likelihood and noise-marginalized optimal statistic. 
These results highlight that the noise-marginalized optimal statistic produces substantially fewer false detections than the maximum-likelihood method. This is consistent with the reduced bias for the noise-marginalized optimal statistic found by \citet{2018PhRvD..98d4003V}. However, \citet{2018PhRvD..98d4003V} report that the maximum-likelihood optimal statistic systematically under-estimates the true GWB amplitude. Here, we find that the bias of the maximum-likelihood optimal statistic appears to work in the opposite sense, in that it results in more false detections of spatial correlations. This could be explained by the fact that the maximum-likelihood method places excessive weight on the CRN terms, while the noise-marginalized optimal statistic marginalizes out these biases. In any case, the message from these results is in agreement with previous analyses \citep{2018PhRvD..98d4003V}: the noise-marginalized optimal statistic is a more accurate and robust tool for measuring spatial correlations in the presence of pulsar timing noise.

\begin{figure}
    \centering
    \includegraphics[width=\linewidth]{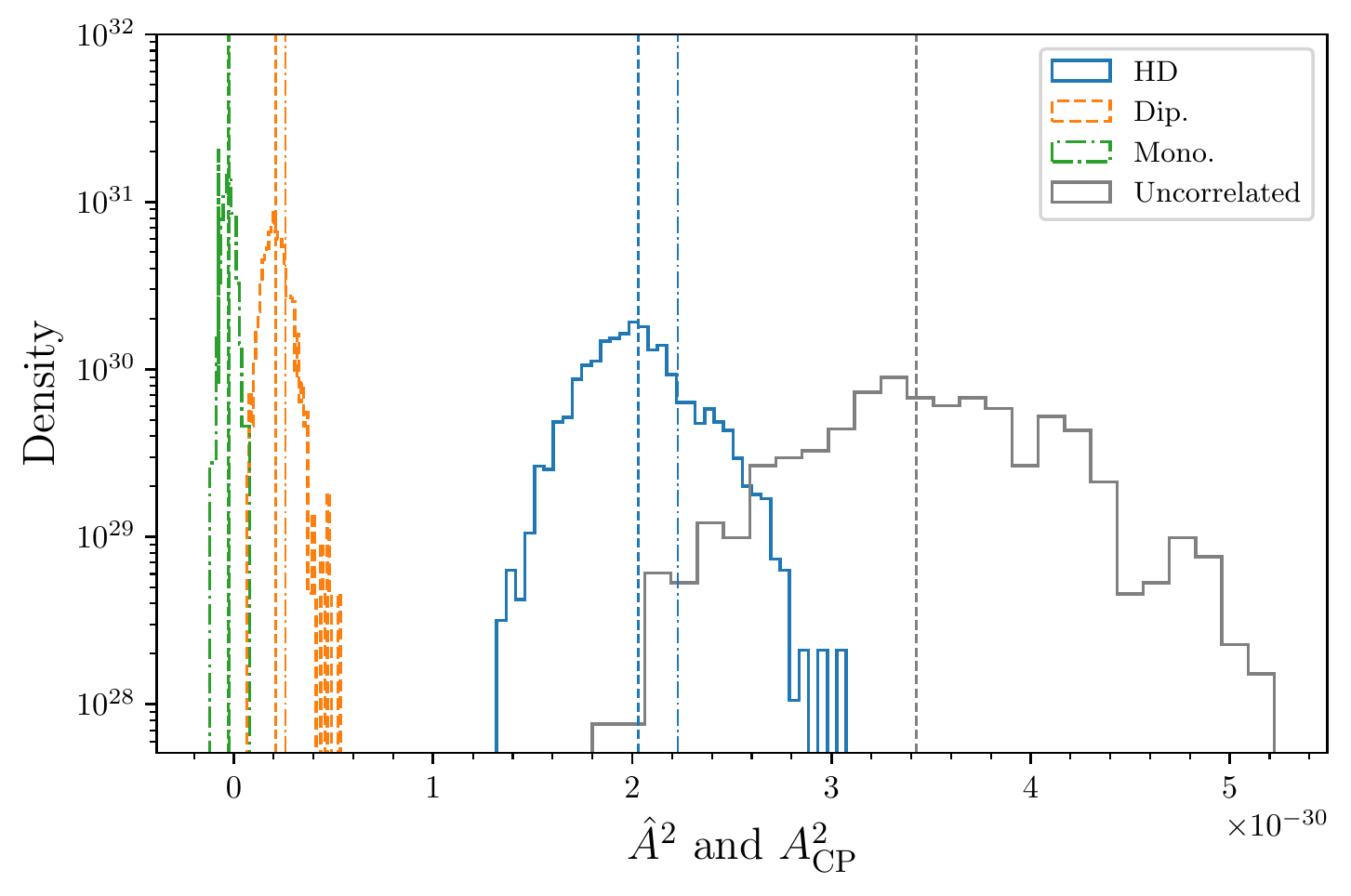}
    \includegraphics[width=\linewidth]{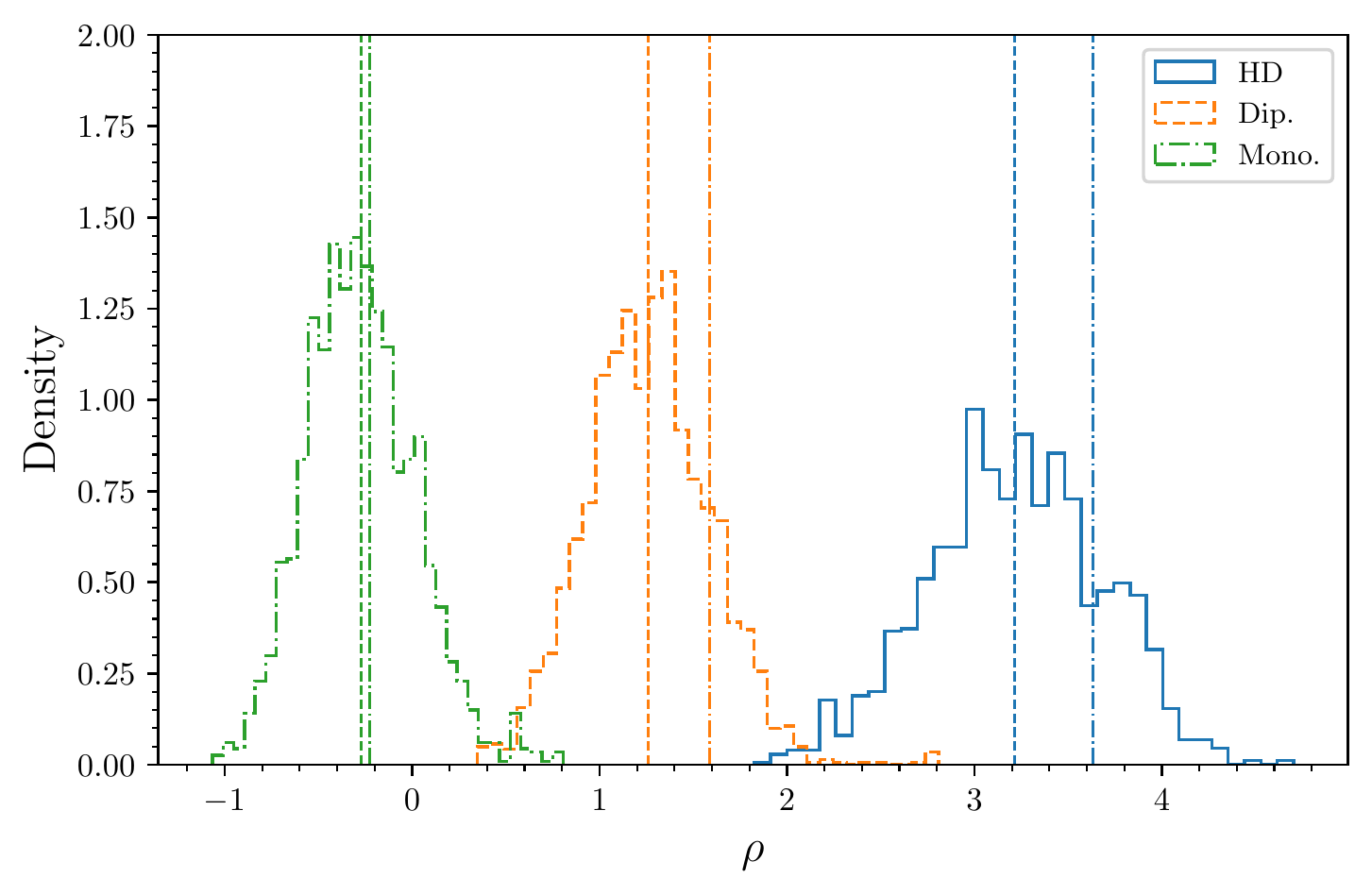}
    \caption{Noise-marginalized optimal statistic (top) and S/N distributions (bottom) for HD, dipolar, and monopolar correlations from a simulation containing only independent pulsar timing noise with $(\Delta \log_{10} P_0, \Delta \gamma) = (1.4, 0.8)$. The squared fixed-slope common noise spectrum amplitude is shown in grey in the top panel. Dashed vertical lines indicate the posterior mean for the corresponding PDF (marginalized over the timing noise parameters), whereas the dash-dotted lines indicate the optimal statistic amplitude and S/N at the maximum-likelihood timing noise parameter values.}
    \label{fig:os_hd_stats}
\end{figure}

\begin{figure}
    \centering
    \includegraphics[width=\linewidth]{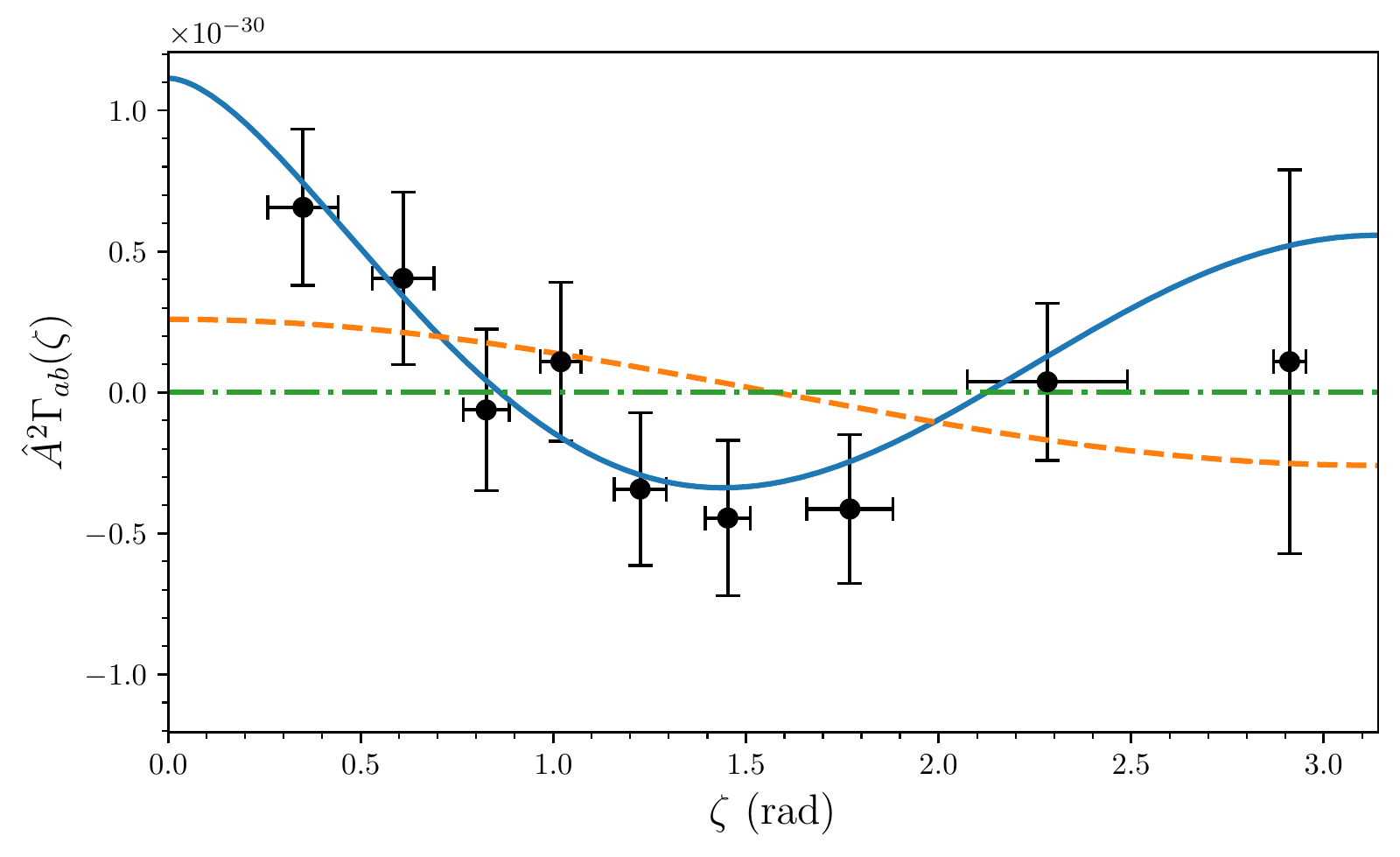}
    \caption{Optimal statistic-derived spatial covariances for a realization with $(\Delta \log_{10} P_0, \Delta \gamma) = (1.4, 0.8)$ having a noise-marginalized S/N of 3.2 for HD correlations. Pulsar pairs are grouped into bins according to their angular offsets before computing the average cross-correlated power per angular bin. The corresponding noise-marginalized optimal statistic and S/N distribution are shown in Figure \ref{fig:os_hd_stats}. The expected covariance for HD, dipole, and monopolar correlations at the level calculated by the optimal statistic are shown in blue, orange, and green, respectively.}
    \label{fig:os_hd_orf}
\end{figure}

The false detection rate is strongly dependent on the choice of maximum-likelihood or noise-marginalized methods, but only moderately dependent on the overlap reduction function being considered. HD correlations appear to have the lowest false detection rate, followed by dipolar and monopolar correlations having higher false detection rates.
A possible reason for this effect is that the simplest ORF in terms of functional form is the monopolar ORF, and a small D.C. offset in the average inter-pulsar covariances could result in a modest signal-to-noise detection of monopolar correlations. On the other hand, dipolar and HD correlations have a more complex spatial signature, which may be more difficult to encounter by chance. We also note that the S/N distributions shown in Figure \ref{fig:os_snr_hist} are positively-skewed for all spatial correlations. In Figure \ref{fig:os_hd_stats}, we show optimal statistic results from a realization with noise-marginalized optimal statistic S/N of $\rho = 3.2$ for HD correlations. In Figure \ref{fig:os_hd_orf}, we show the associated spatial covariance between pulsar pairs, $\hat{A}^{2} \Gamma (\zeta_{ab})$, binned in angular intervals.

\begin{table}
    \centering
    \begin{tabular}{c|cc}
    ORF & Max. Likelihood & Marginalized \\
    \hline
    HD & 0.020 & 0.003\\
    Dipole & 0.023 & 0.004\\
    Monopole & 0.024 & 0.008\\
    \end{tabular}
    \caption{Fraction of realizations with $\rho > 3$ (representing the false detection rate) for different overlap reduction functions (ORFs), and for standard and noise-marginalized optimal statistics.}
    \label{tab:os_ndetections}
\end{table}

% N > 3sigma for hd OS: 0.01977
% N > 3sigma for hd OS: 0.00296
% dp
% N > 3sigma for dipole OS: 0.02290
% N > 3sigma for dipole OS: 0.00439
% mp
% N > 3sigma for monopole OS: 0.02442
% N > 3sigma for monopole OS: 0.00845

\section{Discussion and Conclusions}
\label{sec:Discussion}

Recently, several PTA collaborations have reported detections of a CRN, % \citep{2020ApJ...905L..34A, 2021ApJ...917L..19G, 2021MNRAS.508.4970C, 2022MNRAS.510.4873A}, 
following early indications that such a process may exist \citep{2015MNRAS.453.2576L, 2018ApJ...859...47A}. The detection of a CRN process is consistent with the presence of a GWB \citep{2021PhRvD.103f3027R}, leading to cautious optimism for future GWB detection prospects among the community. However, PTA collaborations also reognize that the detection of a CRN process is not necessarily related to the presence of the GWB in current datasets.

Along this cautionary line, \citet{2021ApJ...917L..19G} raised the possibility of timing noise ``masquerading'' as a CRN. In this work, we have ventured further down this avenue, and have found that a CRN process can be falsely detected under a very wide range of pulsar timing noise conditions, under currently-used hypotheses and methodologies. We have shown that these spurious detections are highly sensitive to the span of the priors on timing noise parameters: when the prior distributions more closely match the real distribution of timing noise parameters, support for a CRN diminishes. While there have been previous and ongoing efforts to account for individual pulsar noise terms \citep[e.g.][]{2021MNRAS.502..478G}, our results are not surprising given the degeneracy between pulsar timing noise and autocorrelation terms of the GWB \citep{2020ApJ...905L...6H,2015MNRAS.449.3293L,2011MNRAS.418..561C, 2010ApJ...725.1607S}. The overwhelming false support for the presence of a CRN under disparate timing noise conditions, under standard assumptions, choices of priors, and models, should lower confidence in the interpretation of recent CRN detections as ``pre-cursor'' detections of the GWB.

On the other hand, while spatial correlations remain undetected, our analysis shows that the detection of spatial correlations (unsurprisingly) is much more robust evidence for the existence of the GWB than the CRN is, in terms of false-detection rates. Our analysis also provides further support for the superior performance of the noise-marginalized optimal statistic over the standard maximum-likelihood method for mitigating biases in the search for spatial correlations.

There have been recent and ongoing efforts to characterise the CRN more robustly and efficiently \citep[e.g.][]{2022ApJ...932L..22G, 2022PhRvD.105h4049T, 2022ApJ...932..105J,2020ApJ...905L...6H}. These works are crucial as PTA collaborations move toward obtaining the first positive detections of the nanohertz GWB. Our results here sound a strong cautionary note of the perverse influence that independent pulsar noise terms and choice of priors can have in the efforts to detect the GWB. Further development is required if we are indeed in the ``intermediate'' S/N regime of GWB detection \citep{2021PhRvD.103f3027R}.
% \todo{Finalise closing remarks here, including links to ongoing work (Boris, etc.)}

% We have shown that the presence of a CRN is overwhelmingly preferred in the presence of wide variations in intrinsic pulsar timing noise. 

% - discuss implication of common noise finding 

% - message - we can't really believe CRN results. Spatial correlations are the way to go, and the noise-marginalized optimal statistic provides a less-biased means of estimating spatial correlations. 
% -- Although a spectrally-common noise process is expected, and consistent with the presence of a GWB, detection of common red noise with current methods/hypotheses is basically meaningless for GWB searches unless you have spatial correlations.

% \section{Conclusions}
% I think the previous section basically acts as a conclusion section?
% \label{sec:conclusions}

\section*{Acknowledgements}
We would like to acknowledge Bill Coles, Paul Baker, Jeff Hazboun, Nihan Pol, and the IPTA gravitational-wave analysis working group for useful discussion during the development of this project. We thank the anonymous reviewer for their constructive comments, and Valentina di Marco for reviewing the text. This work has been carried out by the Parkes Pulsar Timing Array, which is part of the International Pulsar Timing Array.
The Parkes radio telescope (Murriyang) is part of the Australia
Telescope, which is funded by the Commonwealth Government for operation as a National Facility managed by CSIRO. We acknowledge the Wiradjuri people as the traditional owners of the Parkes observatory site. Parts of this research were carried out on the traditional lands of the Wallumettagal people.
This paper includes archived data obtained through the CSIRO
Data Access Portal (\url{https://data.csiro.au}). 
Parts of this research were conducted by the Australian Research Council (ARC) Centre of Excellence for Gravitational Wave Discovery (OzGrav), through project number CE170100004. 
R.M.S. acknowledges support through ARC future fellowship FT190100155. 
BG is supported by the Italian Ministry of Education, University and Research within the PRIN 2017 Research Program Framework, n. 2017SYRTCN. 
Work at NRL is supported by NASA. 
This work made use of the following software: \textsc{astropy} \citep{astropy:2013, astropy:2018}, \textsc{enterprise} \citep{2019ascl.soft12015E}, \textsc{libstempo}~\citep{2020ascl.soft02017V}, \textsc{matplotlib} \citep{Hunter:2007},
\textsc{numpy} \citep{2020Natur.585..357H}, \textsc{ptmcmcsampler} \citep{2017zndo...1037579E}, \textsc{PTASimulate} (\url{https://bitbucket.org/psrsoft/ptasimulate}), \textsc{tempo2} \citep{2006MNRAS.372.1549E}.

%%%%%%%%%%%%%%%%%%%%%%%%%%%%%%%%%%%%%%%%%%%%%%%%%%
\section*{Data Availability}
Simulated datasets and code for this work are available at \url{https://github.com/andrewzic/gwb_crn_sims}.



% \bibliographystyle{mnras}
% \bibliography{bibliography,others} 

%\begin{thebibliography}{99}

%%%%%%%%%%%%%%%%%%%%%%%%%%%%%%%%%%%%%%%%%%%%%%%%%%

%%%%%%%%%%%%%%%%% APPENDICES %%%%%%%%%%%%%%%%%%%%%

% \appendix

% \section{Some extra material}

% If you want to present additional material which would interrupt the flow of the main paper,
% it can be placed in an Appendix which appears after the list of references.

%%%%%%%%%%%%%%%%%%%%%%%%%%%%%%%%%%%%%%%%%%%%%%%%%%

% Don't change these lines
\bsp	% typesetting comment
\label{lastpage}
\end{document}